%% file: main.tex
\documentclass[conference,compsoc]{IEEEtran}
\usepackage{epsfig,endnotes,url}
\usepackage{tcolorbox}
\usepackage{algorithm}
\usepackage{algpseudocode}
\usepackage{amsmath}
\usepackage{graphicx}
\usepackage{subcaption}
\usepackage{cite}
\usepackage{pgfplots}
\usepackage{tikz,array}
\usepackage{soul,xcolor}
\newcolumntype{L}[1]{>{\raggedright\let\newline\\\arraybackslash\hspace{0pt}}m{#1}}
\newcolumntype{C}[1]{>{\centering\let\newline\\\arraybackslash\hspace{0pt}}m{#1}}
\newcolumntype{R}[1]{>{\raggedleft\let\newline\\\arraybackslash\hspace{0pt}}m{#1}}

\newcommand{\eg}{\textit{e}.\textit{g}.,~}

\begin{document}

%don't want date printed
\date{}

%make title bold and 14 pt font (Latex default is non-bold, 16 pt)
\title{\Large \bf I know What You \textit{Sync}: Covert and Side Channel Attacks on File Systems via \textit{syncfs}}

% \author{Anonymous author(s)}
\author{
\IEEEauthorblockN{Cheng Gu}
\IEEEauthorblockA{Department of Computer \\ Science and Engineering \\
University of California, Riverside \\
cgu024@ucr.edu}
\and
\IEEEauthorblockN{Yicheng Zhang}
\IEEEauthorblockA{Department of Electrical \\ and Computer Engineering \\
University of California, Riverside
\\yzhan846@ucr.edu}
\and
\IEEEauthorblockN{Nael Abu-Ghazaleh}
\IEEEauthorblockA{Department of Computer \\ Science and Engineering \\
University of California, Riverside\\
nael@cs.ucr.edu}
}

\maketitle

% Use the following at camera-ready time to suppress page numbers.
% Comment it out when you first submit the paper for review.
\thispagestyle{empty}

\input{abstract}

\input{introduction}

\input{background}

% \input{attackmodel}

\input{reverse_engineering}

\input{covert_channel}

\input{side_channel}

\input{cross_container}

\input{mitigation}

\input{relatedwork}

\input{conclusion}

\bibliographystyle{IEEEtran}
\bibliography{main}

\input{meta_review}

% \theendnotes

\end{document}

%% file: abstract.tex
\begin{abstract}
Operating Systems enforce logical isolation using abstractions such as processes, containers, and isolation technologies to protect a system from malicious or buggy code.  In this paper, we show new types of side channels through the file system that break this logical isolation.  The file system plays a critical role in the operating system, managing all I/O activities between the application layer and the physical storage device. 
%Prior research has identified I/O side-channel vulnerabilities at both the physical and system levels. Early studies demonstrated that I/O pins can serve as leakage vectors for physical side channels, while others showed that contention on I/O can also lead to leakage. However, these attacks are often constrained by the need for physical access or specific application designs. 
We observe that the file system implementation is shared, leading to timing leakage when using common I/O system calls. Specifically, we found that modern operating systems take advantage of any flush operation (which saves cached blocks in memory to the SSD or disk) to flush all of the I/O buffers, even those used by other isolation domains.   Thus, by measuring the delay of \textit{syncfs}, the attacker can infer the I/O behavior of victim programs.  We then demonstrate a \textit{syncfs} covert channel attack on multiple file systems, including both Linux native file systems and the Windows file system, achieving a maximum bandwidth of 5 Kbps with an error rate of 0.15\% on Linux and 7.6 Kbps with an error rate of 1.9\% on Windows. 
In addition, we construct three side-channel attacks targeting both Linux and Android devices. On Linux devices, we implement a website fingerprinting attack and a video fingerprinting attack by tracking the write patterns of temporary buffering files. On Android devices, we design an application fingerprinting attack that leaks application write patterns during boot-up.  The attacks achieve over 90\% F1 score, precision, and recall. Finally, we demonstrate that these attacks can be exploited across containers implementing a container detection technique and a cross-container covert channel attack.

% Additionally, we construct cross-container attacks with \textit{syncfs} including a container detection technique and a cross-container covert channel attack.
% Besides that, we also build three side channel attacks on both Linux and Android devices. On Linux devices, we construct a website fingerprinting attack and a video fingerprinting attack by leaking the write pattern of temporary files. On Android devices, we construct an application fingerprinting attack by leaking the application write pattern during boot-up. The performance of these attacks is above 90\% for F1, precision, and recall.

\end{abstract}

%% file: introduction.tex
\section{Introduction}
\label{sec:intro}

Security and privacy issues in multi-user operating systems (OS) and application software have drawn significant attention from both academia and industry over the past decades. These systems enable concurrent access by multiple users, facilitating efficient sharing of computing resources at both hardware and software levels. At the hardware level, research has identified a series of side-channel attacks targeting shared CPU caches~\cite{rauscher2024idleleak, wang2019papp, liu2015last}, hardware sensors (\eg accelerometers, camera and gyroscope)~\cite{slocum2023going, templeman2012placeraider, gao2022inertiear}, DRAM~\cite{pessl2016drama,van2023dramaqueen}, interconnects~\cite{tan2021invisible,zhang2024beyond}, GPUs~\cite{naghibijouybari2018rendered,wei2020leaky}, and more. On the software side, numerous studies have revealed security vulnerabilities in public OS-provided APIs, leading to OS-level side-channel attacks on systems like iOS~\cite{zhang2018level, wang2023danger}, Android~\cite{diao2016no, zhang2023s}, and Linux~\cite{chen2023sync+, jiang2024sync+}. 

This paper explores OS-level side channels within file systems. File system synchronization is a critical function in the OS, ensuring that data is consistently written from volatile memory (\eg DRAM) to stable storage (\eg SSD or disk drive). This process is essential for data integrity, as it guarantees that recent changes are reliably saved, reducing data loss in the event of a system crash or power failure. Synchronization mechanisms are also integral to many OS optimizations, which improve system performance and resilience. To boost performance, techniques such as batching, scheduling, buffering, and caching are widely implemented~\cite{arpaci2018operating,mogul1994better,seltzer1990disk,silvers2000ubc}. To enhance durability against potential failures, mechanisms such as journaling, write-ahead logging, and copy-on-write are employed~\cite{chidambaram2013optimistic,chidambaram2012consistency,ganger1994metadata,hagmann1987reimplementing}.

As storage device capacities continue to increase into 10s of terabytes or more, it is increasingly common for multiple programs to share the same disk, whether on local machines or cloud-based servers~\cite{gao2019houdini, wang2012disk, bessani2014scfs}. File system synchronization methods like \textit{syncfs} play a crucial role in efficiently and reliably managing data in such shared environments. Unlike the more granular \textit{fsync}, which flushes changes for individual files, \textit{syncfs} operates at the file system level, flushing all pending changes across multiple files in a single call. This approach simplifies synchronization and enhances performance by reducing the overhead associated with multiple \textit{fsync} calls. While these optimizations enhance performance and durability, they also create new avenues for information leakage. 

Recent work~\cite{chen2023sync+, jiang2024sync+} proposed covert channels leveraging contention in \textit{fsync} calls on shared persistent storage. In contrast, our work identifies an orthogonal leakage source by exploiting delay patterns in \textit{syncfs}.  The type of leakage exposed by \textit{syncfs} is fundamentally different from contention-based leakages in that the leakage depends on the write behavior of the victim. 
Thus, unlike \textit{fsync}-based methods, our approach does not rely on the victim application executing any \textit{fsync} calls; the victim simply writes data to the file system.  This leakage is significantly richer than contention leakage, enabling both robust covert channels and high-resolution side-channel attacks. In particular, our threat model assumes a malicious application or process without special privileges, aiming to infer sensitive information about other applications sharing the file system. When the attacker invokes \textit{syncfs}, it flushes not only 
% \cheng{maybe just dirty pages, dirty page cache doesn't sound correct}
its own dirty pages but also the pending changes from other applications. By analyzing the timing delays of \textit{syncfs}, the attacker infers the I/O patterns of these processes. For example, \textit{syncfs} delays increase when other applications perform more writes to the unflushed page caches, such as during webpage loading, video streaming, or the launch of an Android application.

In this work, we first identify two novel leakage vectors through the \textit{syncfs} system call (illustrated in §~\ref{sec:reverse_engineering}). First, we demonstrate how \textit{syncfs} can profile I/O system call footprints, revealing the impact of various I/O operations on execution delays. Second, we show that \textit{syncfs} leakages can infer file write sizes, with latency patterns varying linearly below 4 KB and exhibiting dynamic behavior above 4 KB due to kernel optimizations. 

We then build up a covert channel attack by leveraging \textit{syncfs}-based leakage vectors (illustrated in §~\ref{sec:covert_channel}). By exploiting delay patterns of \textit{syncfs}, the attacker encodes information into binary bits, with high delays representing "1" and low delays representing "0". The synchronization mechanism relies on the Time Stamp Counter (TSC) for precise timing coordination between the sender and receiver. Experiments on various Linux file systems (\eg ext4, ext2. and xfs) and the Windows NTFS file system reveal bandwidths up to 7.61 Kbps with error rates ranging from 0.01\% to 1.9\%.

We also exploit leakage vectors of \textit{syncfs} to develop three end-to-end side-channel attacks (illustrated in §~\ref{sec:side_channel}). These attacks infer sensitive user activities, such as website visits, video streams, and application launches, on Linux and Android devices. By analyzing the timing patterns of \textit{syncfs}, our attacks reveal distinct footprints for these activities. To enhance the effectiveness of our attacks, we employ the Short-Time Fourier Transform (STFT) to extract frequency-domain features from the timing data. These features are further processed using a Convolutional Neural Network (CNN) model. For website fingerprinting, we identify unique \textit{syncfs} delay patterns caused by file writes during website loading, achieving an F1 score of over 93\% in both closed- and open-world scenarios. Similarly, video fingerprinting leverages timing differences from buffering operations, demonstrating high accuracy across streaming platforms like YouTube and Bilibili. In closed-world evaluations, our approach achieves F1 scores exceeding 95\% for YouTube and 89\% for Bilibili. Even in open-world settings, our method maintains robust performance, achieving F1 scores of 92.74\% for YouTube and 87.03\% for Bilibili platforms. On Android devices, application fingerprinting exploits write patterns during app launches, achieving an average F1 score of over 93\%. 

In our final attack, we adapt the \textit{syncfs}-based attack to containerized environments, targeting both detection and covert communication. By monitoring \textit{syncfs} delays, our container detection technique identifies mount and unmount operations associated with container startup and shutdown. These operations create distinct delay spikes, enabling precise identification of container activity. Furthermore, we construct a cross-container covert channel where separate containers communicate by encoding data into \textit{syncfs} delay patterns. Despite the added noise and virtualization overhead inherent in container environments, the attack achieves a bandwidth of 0.23 Kbps with an error rate of 2.4\%.

% \yicheng{Highlight attacks on android}

In summary, the contributions of this paper are: 
\begin{itemize}
    \item \textit{New attack vector.} We analyze the \textit{syncfs} system call on Linux and Android, identifying timing leakages that serve as a novel attack vector.
    \item \textit{Covert channel attack.} We design and implement a fast, resilient covert channel leveraging \textit{syncfs}, demonstrating its effectiveness across Linux and Windows file systems.
    \item \textit{End-to-end side channel attacks.} We present three high-accuracy side-channel attacks to infer visited websites, streamed videos, and foreground applications, achieving over 90\% performance in precision, recall, and F1 score.
    \item \textit{Cross container attacks.} We extend the \textit{syncfs} attack to containerized environments, demonstrating both container detection and a cross-container covert channel with practical feasibility.
    
\end{itemize}

%% file: background.tex
\section{Background and Threat Model}
\label{sec:background}

This section overviews the Linux file system as well as the key system calls relevant to I/O operations. Finally, it presents the threat model assumed in the paper.

\subsection{Linux File System}

The Linux file system is a hierarchical framework that organizes, stores and retrieves data on physical storage devices. It acts as a bridge between user applications and the physical device, enabling efficient and reliable data management across Linux systems~\cite{arpaci2018operating}. 

\begin{figure}[t]
    \centering
    \includegraphics[width=0.48\textwidth]{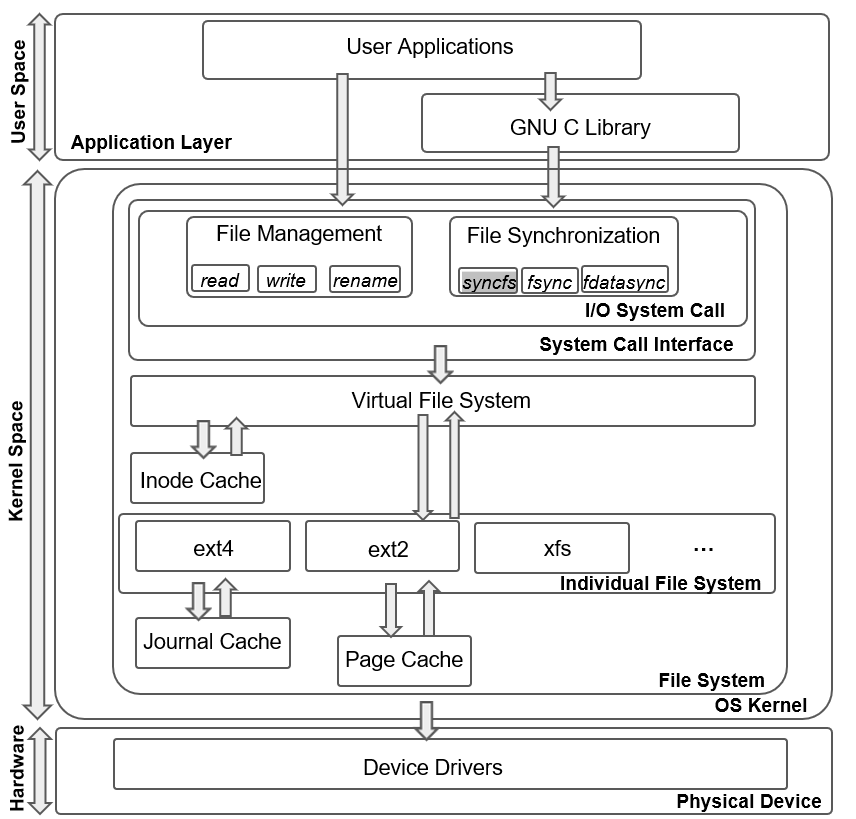}
    % \vspace{-2.0em}
    \caption{Overview of Linux file system and I/O system calls between the application layer and the physical device.}
    \label{fig:file_system_overview}
\end{figure}

Figure~\ref{fig:file_system_overview} shows the structure of the Linux file system, highlighting the flow of I/O system calls from the application layer to the physical device.  The architecture has multiple layers, each with distinct responsibilities. At the top is the \textbf{application layer}, where user applications interact with the file system through libraries like the GNU C Library. These interactions result in system calls, such as \textit{read}, \textit{write}, \textit{rename}, and synchronization calls like \textit{syncfs} or \textit{fsync}. These system calls are the foundation of file operations.

Beneath the application layer, the \textbf{Virtual File System (VFS)} serves as an abstraction layer in the kernel. It standardizes file access, allowing applications to work with files without knowing the details of the underlying file system. This design enables efficient support for multiple file systems, including ext4, XFS, and btrfs, each optimized for different use cases. The VFS bridges the gap between user operations and the specific implementation of individual file systems.

Among these file systems, ext4 is the general-purpose choice and is often set as the default for most Linux installations due to its stability and versatility~\cite{cao2007ext4}. Ext4 provides efficient storage management, large file support, and backward compatibility with earlier file systems like ext2 and ext3. Additionally, its journaling feature enhances reliability by tracking changes before committing them to disk, protecting data in the event of a system crash. For specialized needs, alternative file systems are available. XFS focuses on high-performance workloads with efficient parallel I/O operations, making it ideal for servers and data-intensive applications~\cite{sweeney1996scalability}. Btrfs emphasizes enhanced data integrity with built-in redundancy and snapshots, which are particularly valuable for environments requiring robust data protection~\cite{rodeh2013btrfs}. Finally, NFS supports networked file sharing, enabling seamless access to files across distributed systems~\cite{hitz1994file}. Each of these file systems provides distinct features tailored to specific requirements.

The next layer consists of the actual file systems and their caching mechanisms. These include the \textbf{inode cache}, which stores file metadata, the \textbf{page cache}, which buffers file data, and the \textbf{journal cache}, which tracks transactional logs to ensure data integrity. These caches improve performance by minimizing device access, as we elaborate below.

%\subsection{Linux I/O Buffers} 
% \yicheng{Rewrite it to I/O Buffer, Cheng will rewrite it.}

% Linux I/O buffers include three key caches: \textbf{Inode cache},  \textbf{Page cache.} and \textbf{Journal cache.}.

\textbf{Inode cache.} The inode cache is a memory-resident structure in the main memory that stores recently accessed inodes, which are data structures containing metadata about files and directories. Its purpose is to bypass device reads and writes, improving I/O performance. This caching mechanism plays a critical role in optimizing file system performance.
 
\textbf{Page cache.} The page cache in Linux is an in-memory structure that stores filesystem data to accelerate access. Acting as a buffer, it holds copies of data to enhance system responsiveness and overall performance. When modified, a cached page becomes a "dirty page," requiring eventual write-back to the physical device. The page cache can utilize all available memory, but it dynamically releases memory when needed by other processes. A background process handles the write-back mechanism for dirty pages. Additionally, user applications can trigger write-backs using I/O system calls such as \textit{fsync}, \textit{sync}, \textit{syncfs}.

% The Page cache in the Linux system is an in-memory cache that stores data from the filesystem to speed up data access. It acts as a buffer that holds copies of data and improves the overall system performance and responsiveness. When a page is modified, it becomes a dirty page that needs to be written back to the physical disk. The page cache can use all available memory to cache data. However, if other processes need more memory, a background process would trigger the write-back mechanism that returns dirty pages to the disk. Besides that, user applications can use I/O system calls such as \textit{fsync}, \textit{sync}, \textit{syncfs} to trigger the write-back.    

\textbf{Journal cache.} The journal cache is a dedicated memory area used by journaling file systems to temporarily store updates to the file system's journal before committing them to disk. This mechanism enhances reliability and performance by capturing file system changes in memory. During file operations, the file system logs transaction metadata in the journal cache. Journaling typically follows one of two approaches: metadata journaling, which logs only file system metadata, and full journaling, which records both data and metadata.

% The journal cache is a memory area used by filesystems to store updates to filesystem's journal before committing to the physical disk. Journaling filesystems utilize this cache to improve reliability and performance by temporarily holding filesystem changes. When a file operation happens, the filesystem records metadata about that transaction to the journal cache. Filesystems usually have two types of journaling: metadata journaling that only writes filesystem metadata to journal cache, and full journaling that writes both data and metadata to journal cache.

Finally, at the bottom of the stack, \textbf{device drivers} handle communication with the physical device. They execute the low-level commands needed to complete read and write requests.

\subsection{System Call for I/O}

\label{subsec:System_call}
\textbf{File management system calls.} System calls for I/O are essential functions that allow user programs to interact with the kernel to perform I/O operations. Common I/O system calls include \textit{open}, \textit{read}, and \textit{write} to transfer data between the disk and the program memories. Applications commonly use these I/O system calls due to its functionality and low overhead.

\textbf{File synchronization system calls.} Synchronization system calls allow manual synchronization of pending changes in I/O buffers. Commonly used calls include \textit{fsync} and \textit{fdatasync}, which flush dirty pages from I/O buffers associated with individual files. In this paper, we focus on the Linux system call \textit{syncfs}. Unlike other synchronization calls, \textit{syncfs} operates at the file system level. It flushes all dirty pages related to the file's superblock, synchronizing every file within the disk partition to the physical device.

\textbf{\textit{syncfs} workflow.} \textit{syncfs} consists of three main components: synchronizing inodes, synchronizing data, and performing file system-specific operations. Each component uses a lightweight function to initiate the write-back process, followed by a blocking function that ensures the operations are complete.

For inode synchronization, the function \texttt{writeback\_inode\_sb()} marks inodes as dirty and initiates write-back without waiting, while \texttt{sync\_inode\_sb()} ensures all dirty inodes are flushed to disk. This approach extends to data synchronization as well.

The function \texttt{sync\_blockdev\_nowait()} sends write-back requests and returns immediately. At the end of \textit{syncfs}, \texttt{sync\_blockdev()} is called in a loop to flush all dirty pages, incorporating an I/O throttling mechanism to periodically put the process to sleep. This throttling prevents the physical device from becoming overwhelmed and avoids saturating the I/O bandwidth. 

File system-specific operations depend on the implementation of each file system. File system-specific operations vary based on the implementation of each file system. A representative example is \texttt{sb->s\_op->sync\_fs()}, which handles synchronization tasks unique to the file system's design. In ext4, this operation records metadata to the journal to maintain data consistency. Finally, \texttt{blkdev\_issue\_flush()} flushes the physical device's internal cache, guaranteeing that the journal log is physically written to the disk.

% \textit{syncfs} contains three main parts, synchronizing inodes, synchronizing data, and file system specific operations. Within each part, a lightweight function is called to initiate the write-back process followed by the same function but waits for the operations to end. When synchronizing inodes, \textit{writeback\_inode\_sb()} marks the inodes dirty for write-back and returns without waiting, and \textit{sync\_inode\_sb()} ensures all dity inodes are flushed back to the disk. This idea is also applied to synchronize data. \textit{sync\_blockdev\_nowait()} sends the write-back request and returns without waiting. \textit{sync\_blockdev()} is called at the end of \textit{sybcfs} and does write-back in a loop to flush all the dirty pages. Inside that loop, an I/O throttling mechanism is called to put the process in sleep periodically. This I/O throtting mechanism prevents overwhelming the physical disk and saturating the I/O bandwidth. File system specific operations(\textit{sb->s\_op->sync\_fs()}) may vary based on different implementations in each file system. In ext4, this operation records only the metadata to the journal to ensures data consistency. Then, \textit{blkdev\_issue\_flush()} is called to flush the physical disk's internal cache. The journal log is now guaranteed to be physically written to the disk.

\subsection{Threat Model}
\label{sec:threatmodel}

% \yicheng{Needs a figure for threat model. Highlight the shared file system.}

Our threat model assumes that a \textit{spy} application and a \textit{victim} application execute on the same shared file system. In Linux or Windows, it is common for all applications to operate within the same file system, especially in standard desktop and server installations. We do not assume the spy has superuser privileges or access to specialized hardware. The spy operates as a regular user and periodically invokes the \textit{syncfs} system call, which requires no root permissions. The only requirement is a Linux kernel version 2.6.39 or higher, or an Android OS version 9 or later, both of which support the \textit{syncfs} system call.
% \cheng{Android requirement added}
%\cheng{do we also add the requirement for android?} \yicheng{Yes, please add it as well.}

The spy initiates the attack by repeatedly calling \textit{syncfs} in a loop to clear the page cache as preparation. It continuously calls \textit{syncfs} in the background while recording the system call's time delay. When the victim process runs, any I/O system calls it performs that modify data or metadata increase the delay observed by the spy. The spy analyzes this delay pattern to infer the victim's file system activities.

Our attacks generalize to most file systems, including both native and non-native file systems on Linux and Android. Variations in the implementation of file system-specific operations within \textit{syncfs} may cause its performance to differ across file systems.
In the covert channel attack, the sender performs distinct file system operations to encode bits "0" and "1". The receiver continuously invokes \textit{syncfs} in the background to detect these signals. By measuring the differences in \textit{syncfs} delays, the receiver decodes the transmitted message.
In the side-channel attack, the attacker flushes the page cache at the file system level to monitor the victim's I/O operations. For fingerprinting attacks, the attacker captures the victim's write patterns and leverages them to classify user activities. Specifically, in the Android application fingerprinting attack, the attacker records the I/O activities of applications during boot-up to identify them.

% Our attacks can also be generalized to all file systems including both native and non-native file systems on Linux and Android. Different implementations in the file system specific operation in \textit{syncfs} may casue the performance of \textit{syncfs} vary. In our covert channel attack, the sender conducts different file system operations to represent 0 and 1. The receiver keeps calling \textit{syncfs} in the background to capture these signals. Based on the difference in delays of these \textit{syncfs}, the receiver can recover the message. In the side channel attack, the attacker keeps flushing the page cache in the file system scale to detect the victim's I/O operations. In our fingerprinting attacks, the write patterns of the victim processes are captured by the attacker and used to classify the user activities. In the Android application fingerprinting attack, the I/O activities of applications during boot up are captured by the attacker. 

%% file: reverse_engineering.tex
\section{Leakage Vectors through \textit{syncfs}}
\label{sec:reverse_engineering}

In this section, we introduce a \textit{new} attack vector on the Linux file system via \textit{syncfs}. We demonstrate how \textit{syncfs} can be leveraged to profile the footprints of other Linux I/O system calls. Additionally, we focus on the \textit{write} system call and show that the written file size can be inferred using \textit{syncfs} leakages.

\subsection{Measuring I/O System Call Footprint via \textit{syncfs}}

The \textit{syncfs} system call operates at the file system level rather than on individual files. It ensures that all dirty data and metadata in the file system are flushed to the storage device. In ext4, \textit{syncfs} flushes \textbf{dirty inodes, dirty pages, and the journal}. We observed that the execution time of \textit{syncfs} increases when the file system has pending changes in these I/O buffers caused by other operations (\eg \textit{write} and \textit{ftruncate}). 

\textbf{Experiment 1: I/O operations delay \textit{syncfs} execution.} Our first experiment tests the delay impact on \textit{syncfs} caused by different I/O operations. We selected four operations that introduce delays in \textit{syncfs}: \textit{write}, \textit{ftruncate}, \textit{write(O\_SYNC)}, and \textit{rename}. These operations affect different I/O buffers, as summarized in Table~\ref{tb:latency_measurement}:
\begin{itemize}
    \item Operation \textit{write} appends or updates data asynchronously to a file, affecting the page cache, journal cache, and inode cache.
    \item Operation \textit{write(O\_SYNC)} writes data synchronously, affecting only journal cache.
    \item Operation \textit{ftruncate} adjusts the file size (truncate or extend), impacting the inode cache and journal cache.
    \item Operation \textit{rename} changes a file’s name, modifying the journal cache and inode cache.

\end{itemize}

\textbf{{Experiment platform.}} We use an NVIDIA Jetson AGX Orin machine with the ext4 file system set to default and a kernel version of 5.15.136-tegra. To measure timing, we use the ARM Generic Timer counter register, CNTVCT\_EL0~\cite{arm_ctr}. We execute each I/O operation followed by a \textit{syncfs} call to measure its impact on delay. For \textit{write}, we write a full data block (4 KB) to a file, adding dirty pages and inodes to I/O buffers. The \textit{syncfs} delay includes the extra time to flush the page cache, inode cache, and journal log. 
To test \textit{ftruncate} and \textit{rename}, we shrink a file size to 0 and rename a file, respectively. These two operations modify only metadata, adding dirty inodes to the buffer. The \textit{syncfs} delay reflects flushing the inode cache and journal log.  For \textit{write(O\_SYNC)}, we set the \texttt{O\_SYNC} flag and write a full data block (4 KB) synchronously. Since data and metadata are directly written, \textit{syncfs} only needs to flush the journal log. Additionally, we measure the base delay of \textit{syncfs} as the baseline, representing its execution time without any I/O processes. 

\textbf{Results.} We measure the latency of each operation 1,000 times and calculate the average and standard deviation. Table~\ref{tb:latency_measurement} summarizes the results. Compared to the baseline, all four operations increase the \textit{syncfs} latency by at least 16 times. Among them, \textit{write} introduces the highest latency, averaging 121,091 clock cycles, as it affects three buffers: the page cache, journal, and inode cache. 

In contrast, \textit{write(O\_SYNC)}, which only updates the journal log, incurs the lowest latency, averaging 41,406 clock cycles. Operations \textit{ftruncate} and \textit{rename} both modify the journal and inode cache, resulting in average latencies of 61,315 and 66,774 clock cycles, respectively.

\begin{table}[tb]
\caption{\textit{syncfs} leakages for different I/O operations.}
% \small
\begin{tabular}{|L{2.2cm}|C{2.0cm}|C{1.2cm}|C{1.5cm}|}

\hline
\textbf{I/O operation}         & \textbf{Affected buffers}              & \textbf{Average latency (cycles)} & \textbf{Standard deviation} \\ \hline \hline
baseline  & N/A                                    & 2509                           & 491                      \\ \hline
write                          & Page cache, journal and inode   & 121092                        & 11436                    \\ \hline
write(O\_SYNC)                 & journal                          & 41406                          & 4670                      \\ \hline
ftruncate                      & journal and inode              & 61315                          & 6916                     \\ \hline
rename                         & journal and inode           & 66774                          & 8134                     \\ \hline
\end{tabular}
\label{tb:latency_measurement}

\end{table}

\begin{tcolorbox}[colback=gray!10,colframe=black,boxrule=0.5pt,arc=2mm,outer arc=2mm,]
\textit{\textbf{Observation 1}: I/O operations delay \textbf{syncfs} execution time, with different operations causing varying impacts on \textbf{syncfs} latency.
}
\end{tcolorbox}

\textbf{Experiment 2: impact of concurrent I/O operations on \textit{syncfs} latency.} Our first experiment analyzed the impact of a single I/O operation on \textit{syncfs} latency. However, in real-world scenarios, multiple I/O operations are more likely to be executed concurrently to handle multiple files. 
% \cheng{I think it is better to mention it is hard to flush only one file as our motivation for this.} \yicheng{how about this?}
%\cheng{looks good}
For example, several files might be renamed, truncated, or written simultaneously. Thus, our second experiment measures the latency impact on \textit{syncfs} caused by concurrent I/O operations.

\textbf{Experiment platform.} This experiment is conducted on a Linux PC with an Intel Xeon E5-2698v4 CPU running Ubuntu 20.04. We use the unprivileged \textit{rdtsc} timer to measure latency on the ext4 file system. The tested I/O operations in this experiment include: \textit{write(64B)}, \textit{write(O\_SYNC)(64B)}, and \textit{ftruncate(0)}, as detailed below:

\begin{itemize}
    \item Operation \textit{write(64B)}: Writes 64 bytes of data to a file using the standard \textit{write()} system call.
    \item Operation \textit{write(O\_SYNC)(64B)}: Writes 64 bytes of data directly to the disk, bypassing the OS's page cache.
    \item Operation \textit{ftruncate(0)}: Truncate a file of 64 bytes to 0 bytes. 
\end{itemize}

\begin{figure}[tb]
    \centering
    \includegraphics[width=0.45\textwidth]{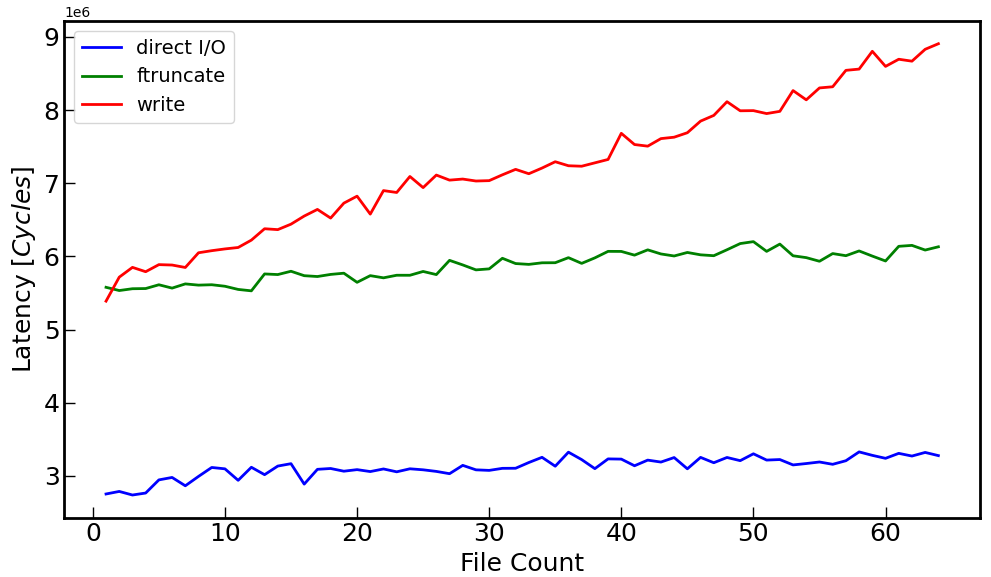}
    % \vspace{-2.0em}
    \caption{\textit{syncfs} delay of concurrent I/O system calls on ext4}
    \label{fig:re_concrrent_IO}
\end{figure}

\begin{figure*}[ht]
    \centering
    \includegraphics[width=\textwidth]{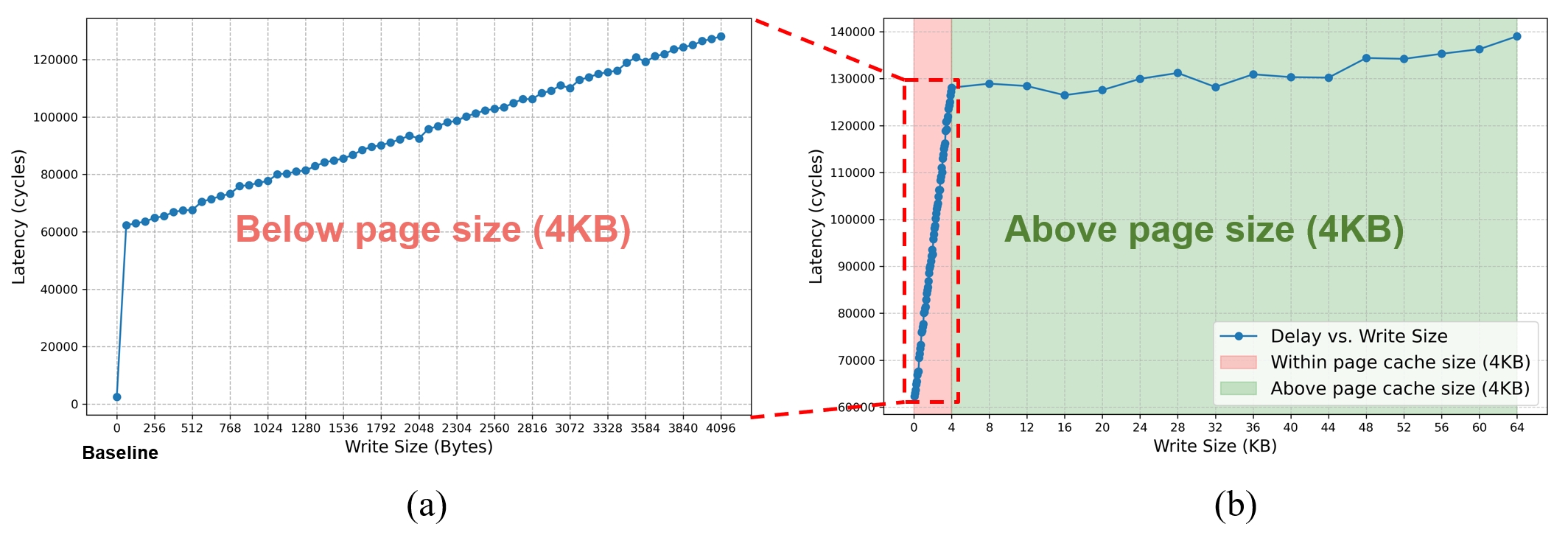}
    % \vspace{-1pt}
    \caption{\textit{syncfs} latency with different \textit{write} size: (a) Below page size (4KB), (b) Above page size (4KB).} 
    % \vspace{-1pt}
    \label{fig:write_size}
\end{figure*}

In this experiment, we measure the \textit{syncfs} delay caused by flushing extra dirty buffers generated by three I/O system calls while increasing the number of files to process. Figure~\ref{fig:re_concrrent_IO} shows the delay patterns for \textit{write(64B)}, \textit{write(O\_SYNC)(64B)}, and \textit{ftruncate}. We observe that the \textit{syncfs} delay for all three I/O system calls increases as the file count grows, but with varying slopes. The calculated slopes of the delay patterns are 6,163 (\textit{write(O\_SYNC}), 9,626 (\textit{ftruncate}), and 48,612 (\textit{write}), reflecting the differences in overhead caused by each operation.

For direct I/O (\textit{write(O\_SYNC)(64B)}), the \textit{syncfs} latency increases slightly, even with multiple concurrent operations, as this call primarily affects the journal cache. Similarly, the latency of \textit{syncfs} shows a modest increase for concurrent \textit{ftruncate} operations, which generate additional dirty journal and inode entries for flushing. In contrast, the \textit{write} operation, which affects the journal, inode, and page cache, causes a steep linear increase in \textit{syncfs} latency as the number of processed files increases. Additionally, extra inode entries, journal updates, or dirty pages cannot be merged by the I/O scheduler. As the scheduler serializes the write-back requests, the resulting overhead leads to progressively longer delays for \textit{syncfs}.
% \cheng{I want to mention some details about the overhead that causes increasing delay. Additional inode entries, journal, or dirty pages can not be merged by the I/O scheduler. As the scheduler serialize the write-back requests, the overhead causes the increasing delay. }

\begin{tcolorbox}[colback=gray!10,colframe=black,boxrule=0.5pt,arc=2mm,outer arc=2mm,]
\textit{\textbf{Observation 2}: An increasing number of I/O operations also increases \textbf{syncfs} delay. Different operations have varying degrees of impact on the growth of \textbf{syncfs} latency.}
\end{tcolorbox}

% \noindent \textbf{Scheduling Overhead.} We also observe that the \textit{syncfs} delay of all three I/O system calls increases as the file count increases. We believe the increase in \textit{syncfs} delay is related to overhead in the I/O scheduler. Each file has its own data structures, such as Inode and journal. Modifying different files marks dirty pages that are not contiguous, which means the scheduler must serialize them when flushing. For instance, \textit{ftruncate} modifies multiple files and marks non-contiguous dirty pages in both the Inode and journal cache. Therefore, the I/O scheduler must decide the order of flushing Inodes and journals in these caches. 

% By computing the slope of the delay pattern of these I/O system calls, we get 6163 (direct I/O), 9626 (ftruncate), and 48612 (write). The slops reflect the difference between the overhead in each I/O system call. For instance, flushing \textit{direct I/O} only involves the journal cache, and its slop is less than \textit{ftruncate} which includes flushing both the inode cache and the journal cache. We also conclude that there is more scheduling overhead in the page cache than the other I/O caches since the slope of \textit{write} is much higher. 

%\noindent \textbf{{Experiment platform.}}  Android and Linux
\subsection{Inferring \textit{write} Size via \textit{syncfs}}
\label{sec:re_2}

\textbf{Motivation.} Among these I/O system calls, \textit{write} is most frequently used in both application and system-level programs. It is essential for writing various outputs. The size parameter in \textit{write} directly specifies the number of bytes to be written from a buffer to a file. It can always reflect the actual size of the object being written to a file.  Therefore, in this experiment, we explore whether \textit{syncfs} can leak the write size of a \textit{write} system call.

% \yicheng{Highlight why we focus write? 1. Most frequently use 2. Write size control the total size of dirty pages.}
 \textbf{Experiment 3: infer write size via \textit{syncfs}.} We have two steps for this experiment: within page size (4KB), and above page cache size. Within the page size, we set the initial write size as 64 bytes. It keeps increasing with a stride of 64 bytes and ends at 4 KB. Above page size, we set the initial write size as 4 KB. It keeps increasing with a stride of 4 KB and ends at 64 KB. The experiment was conducted on an NVIDIA Jetson AGX Orin machine with the ext4 file system and running kernel version 5.15.136-tegra.

 \textbf{Results when below the page size.} Figure~\ref{fig:write_size} (a) shows the zoomed-in \textit{syncfs} latency for write sizes ranging from 64 bytes to 4,096 bytes. The baseline latency, when no write occurs, is approximately 2,000 cycles. However, at a write size of 64 bytes, there is a significant jump in latency to around 62,000 cycles.

As the write size increases within the page size limit (4 KB), the \textit{syncfs} latency grows linearly. This is because larger write sizes generate more dirty page cache entries, journal updates, and inode changes, leading to longer flushing times for \textit{syncfs}. Specifically, \textit{syncfs} performs I/O in two stages: \texttt{sync\_blockdev\_nowait()}, which initiates a write-back request, and \texttt{sync\_blockdev()}, which loops until the write-back completes. As the write size grows, \textit{sync\_blockdev()} takes longer to exit the loop. Within this loop, periodic I/O throttling forces the process to sleep for a fixed duration, further increasing delays. These cumulative delays result in the observed linear relationship between write size and \textit{syncfs} latency.

\textbf{Results when above the page size.} However, when the write size exceeds 4 KB, the increase in \textit{syncfs} latency slows down significantly, as shown in Figure~\ref{fig:write_size} (b). This slowdown is due to parallelism in the I/O process. When the write size exceeds 4 KB, dirty pages are merged and processed by multiple flusher threads~\cite{linux_flusher}. 

We also observe more fluctuations in the latency pattern for write sizes above 4 KB. This is primarily caused by the kernel’s dynamic adjustment of flusher threads, which are responsible for writing dirty pages back to the physical disk~\cite{mario2017low}. The Linux kernel adjusts the number of flusher threads based on factors such as system load and memory pressure. As a result, \textit{syncfs} delay does not consistently increase with larger write sizes above 4 KB. While the fluctuations make the pattern noisy, it is still possible to infer the approximate range of write sizes beyond 4 KB. 
% \yicheng{any reference to dynamic adjustment.}

% \yicheng{flush thread api? add link to this flusher thread? https://github.com/firmianay/Life-long-Learner/blob/master/linux-kernel-development/chapter-16.md} Focus on dynamic allocate flusher threads.

\begin{tcolorbox}[colback=gray!10,colframe=black,boxrule=0.5pt,arc=2mm,outer arc=2mm,]
\textit{\textbf{Observation 3}:  For write sizes below 4 KB, \textit{\textbf{syncfs}} latency grows linearly. For write sizes above 4 KB, \textit{\textbf{syncfs}} latency slows due to I/O parallelism, with fluctuations from dynamic flusher thread adjustments. }
\end{tcolorbox}

% \subsection{Inferring Concurrent I/O System Calls }
% \yicheng{Need to define concurrent I/O system call here.}

% \noindent \textbf{{Motivation.}} Move to subsection of 3.1

 \subsection{Robustness of Side Channel against Noise}
In real-world settings, background applications may run concurrently with an attacker, introducing noise that can interfere with the side-channel. To assess the impact of such interference, we use the signal-to-noise ratio (SNR), a standard metric to evaluate side-channel leakage{~\cite{mangard2004hardware}}. Specifically, we apply the following SNR equation: $SNR = \frac{Var(Signal)}{Var(Noise)}$.
In particular, an SNR value greater than 1 is exploitable as recent work highlighted{~\cite{zhang2023s}}, with higher values indicating a higher quality channel.  We collect the \textit{syncfs} signals of browsing a website while introducing background noise in two ways: (1) opening additional browser tabs and (2) running I/O noise microbenchmarks. The I/O noise microbenchmarks launch multiple threads to continuously generate dirty pages at a rate of 81 MB/s.  We computed the SNR by averaging the results over 10 traces. We observe that the SNR drops from 15.77 (without noise) to 15.54 and 12.78 when one and five browser tabs are opened, respectively. In a different experiment, when one and five I/O noise workloads are introduced to run concurrently with our spy and victim, the SNR decreases to 7.66 and 4.75, respectively. 

% \yicheng{@Cheng, could you add the reference to this microbenchmarks?}

%% file: covert_channel.tex
\section{Covert Channel Attack}
\label{sec:covert_channel}
In this section, we exploit the findings from Section~\ref{sec:reverse_engineering} to build a fast and resilient covert channel. In Section~\ref{subsec:covert_design}, we describe the design of the covert channel and synchronization mechanisms. We evaluate and show results in Section~\ref{subsec:covert_evaluation}.

\subsection{Covert Channel Design}
\label{subsec:covert_design}
In this section, we outline the design of the covert channel. We assume the receiver (spy) and sender (trojan) processes operate on the same file system, sharing the same permission level as other normal processes. 

\begin{figure}[t]
    \centering
    \includegraphics[width=0.4\textwidth]{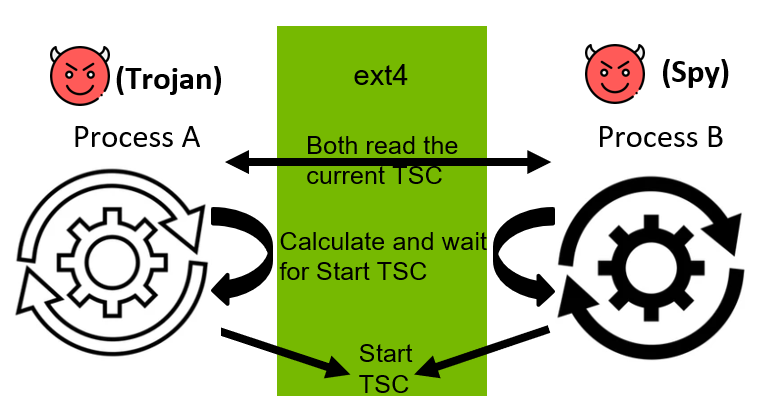}
    % \vspace{-2.0em}
    \caption{The \textit{syncfs} covert channel uses TSC for synchronization on ext4 before transmission.}
    \label{fig:covert_channel_synchronization}
\end{figure}

\textbf{{Covert channel synchronization.}} A key component is the synchronization between the sender and receiver, as it significantly enhances bandwidth and reduces the error rate. We follow the synchronization method proposed in prior work~\cite{rauscher2024idleleak}. As illustrated in Figure~\ref{fig:covert_channel_synchronization}, we utilize the TSC (Time Stamp Counter) to synchronize the sender and receiver. Since TSC is a register timer accessible to any process, the current TSC value is shared between the sender and receiver. The sender and receiver both compute a start time by adding a predefined constant to the current TSC value. They wait until the computed start time, after which the sender begins transmitting the message with a starter code. Once the receiver detects the starter code, the covert channel is established.

\textbf{{Design.}} In the Section~\ref{sec:reverse_engineering}, we have observed that multiple I/O system calls can cause delay in \textit{syncfs}. The communication channel based on \textit{syncfs} leverages the write + flush mechanism in I/O buffers. The sender controls a file and writes data or metadata to it, marking dirty pages in the I/O buffers. To minimize overhead, the sender uses direct I/O (write(O\_SYNC)), which requires only the journal cache to be flushed. The receiver observes a higher \textit{syncfs} delay when flushing the dirty pages. Conversely, when the sender is inactive, the receiver experiences the base \textit{syncfs} delay, which occurs without any I/O operations. 

% \cheng{do we need to mention that the receiver also needs to control a file? } \yicheng{it should be fine. Just make this file is not shared with sender}

The algorithms for the receiver and sender are visualized in Algorithm~\ref{alg:receiver} and \ref{alg:sender}, respectively. Once synchronization via TSC is completed, the receiver begins measuring the delay of \textit{syncfs}. It decodes the observed delay patterns into bit representations, stopping when it detects the ending code, a specific pattern indicating the end of the message. 

We define bit '0' and bit '1' using distinct delay patterns, as shown in Algorithm~\ref{alg:receiver}. A higher delay caused by flushing I/O buffers represents bit '1', while a base delay without I/O represents bit '0'. In our covert channel, the sender adjusts the delay pattern by either writing 64 bytes to a file using direct I/O or executing an NOP loop for a defined number of iterations.

\subsection{Covert Channel Evaluation}
\label{subsec:covert_evaluation}

\begin{algorithm}[t]
\caption{Receiver for Covert Channel} \label{alg:receiver}
\begin{algorithmic}[1]
% \small
\State // curr\_TSC is the current TSC value  
\State // $Pattern_{\text{receiver}}[N]$ is an array of N samples to record the delay pattern
\State // $Message_{\text{receiver}}$ is a vector to store the parsed bits
\State // $fd$ is a file controlled by the receiver
\State // $T$ is the threshold to differentiate bits '1' and '0'

\State Synchronization(curr\_TSC);
\For{$i \gets 0$ \textbf{to} $N-1$}
    \State $start \gets clock();$
    \State $syncfs(fd);$
    \State $end \gets clock();$
    \State // Check for the ending code
\EndFor

\end{algorithmic}
\end{algorithm}

\begin{algorithm}[t]
\caption{Sender for Covert Channel} \label{alg:sender}
\begin{algorithmic}[1]
% \small
\State // $Message_{\text{sender}}[N]$ is an array of $N$ bits used to send a message
\State // $fd$ is a file controlled by the sender
\State // $R$ is the number of iterations to wait for flushing
\State // $K$ is the number of iterations of no operation

\State Synchronization()

\For{$i \gets 0$ \textbf{to} $N-1$}
    \If{$D_{\text{sender}}[i] == 1$}
            \State write(fd,string,size);
            \For{$j \gets 0$ \textbf{to} $R-1$}
                \State NOP;
        \EndFor
    \Else
        \For{$j \gets 0$ \textbf{to} $K-1$}
                \State NOP;
        \EndFor
    \EndIf
\EndFor

\end{algorithmic}
\end{algorithm}

% \label{subsec:covert_evaluation}

We evaluate the \textit{syncfs} covert channel attack with two metrics: bandwidth and error rate. The bandwidth is measured in Kb per second, and the error rate is measured with Levenshtein edit distance~\cite{miller2009levenshtein}.

\begin{figure}[tb]
    \centering
    \includegraphics[width=0.49\textwidth]{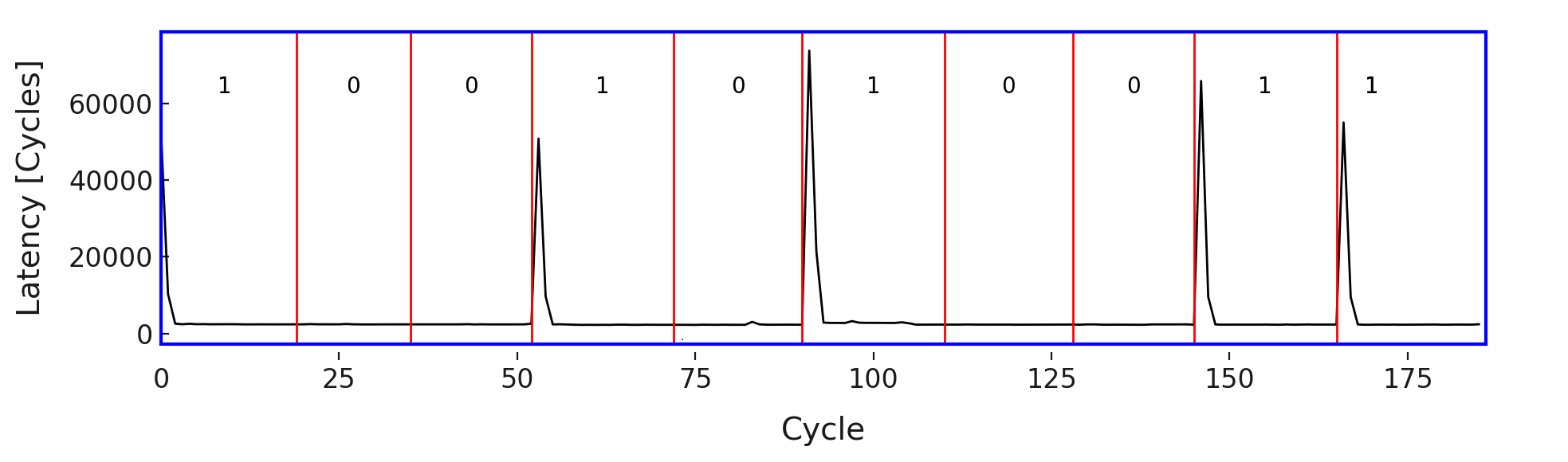}
    % \vspace{-2.0em}
    \caption{The \textit{syncfs} covert channel transmission has a clear difference between bit '0' and '1' on ext4.}
    \label{fig:covert_channel_demo}
\end{figure}

\textbf{Covert channel demonstration.} Figure~\ref{fig:covert_channel_demo} illustrates our covert channel on the ext4 file system, where the distinct latency pattern of \textit{syncfs} encodes data: higher latency represents bit "1," while lower latency represents bit "0."

\textbf{Linux file system.} We implement the covert channel on multiple file systems, including ext4, ext2, and xfs, and evaluate it based on bandwidth and error rate, as summarized in Table~\ref{tb:covert_results}. Due to the higher average latency of \textit{syncfs}, we achieve the lowest bandwidth of 0.94 Kbps with an error rate of 0.01\% on ext4. On ext2, the bandwidth reaches 3.28 Kbps, but with a higher error rate of 1.6\%, attributed to increased background noise. On XFS, we achieve the highest bandwidth of 4.98 Kbps with a moderate error rate of 0.15\%.

\textbf{Windows file system.} We also explore the scenario of running Linux on a Windows machine, such as using WSL, which provides a lightweight Linux environment~\cite{barnes2021pro}. In this setup, the Linux system operates on the NTFS file system via the NTFS3 driver. Implementing the syncfs covert channel on NTFS achieves the highest bandwidth of 7.61 Kbps with a 1.9\% error rate. The increased bandwidth compared to Linux native file systems is due to the significantly lower \textit{syncfs} delay for direct I/O operations.

\begin{table}[tb]
\caption{Covert channel results: Bandwidth (Kbps), and Error rate (\%).}
% \normalsize
% \small
\centering
\begin{tabular}{|l|c|c|} 
\cline{2-3}
%\multicolumn{1}{c|}{\textbf{}} & \textbf{F1}   & \textbf{Prec} & \textbf{Rec}  \\ \hline \hline
\multicolumn{1}{c|}{\textbf{}} & \textbf{Bandwidth} & \textbf{Error rate} \\ \hline  \hline
ext4 &0.94  & 0.01 \\ \hline  
ext2  &3.28  &1.6   \\ \hline  
xfs &4.98 &0.15  \\ \hline 
ntfs &\textbf{7.61}   &1.9  \\ \hline 
\end{tabular}
\label{tb:covert_results}
\end{table}

% \noindent \textbf{Parallel covert channels.} One way to enhance covert channel performance is to run multiple covert channels in parallel when several file systems are mounted on the same machine. Since \textit{syncfs} operates independently at the file system level, flushing multiple file systems concurrently does not introduce additional noise. 

%% file: side_channel.tex
\section{Side Channel Attacks}
\label{sec:side_channel}
In this section, we utilize the leakage vectors identified in Section~\ref{sec:reverse_engineering} to design three side-channel attacks: website fingerprinting, video fingerprinting, and application fingerprinting. We assume a malicious process continually profiles timing side-channel leakage through \textit{syncfs} while a victim process visits a website, watches a video, or launches an application. This approach enables the identification of the specific website, video, or application accessed by the victim. The website and video fingerprinting attacks are demonstrated on Linux file systems (Section~\ref{subsec:web_finger} and Section~\ref{subsec:video_finger}), while the application launch identification attack (Section~\ref{subsec:app_finger}) is conducted on Android devices.

\begin{figure*}[ht]
    \centering
    % First subfigure
    \begin{subfigure}[b]{0.32\textwidth}
        \centering
        \begin{tikzpicture}
            \node[anchor=south west,inner sep=0] (image) at (0,0) {\includegraphics[width=\textwidth]{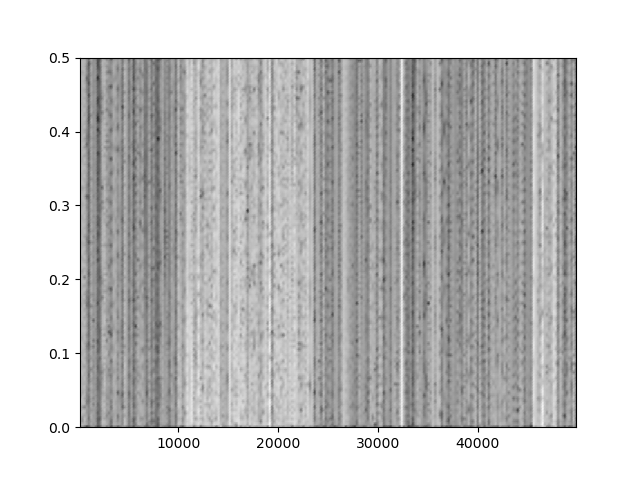}}; % Replace with your actual image file
            \node[above=-5mm] at (image.south) {Sample}; % Move "Sample" slightly farther away
            \node[rotate=90, anchor=south, yshift=-4mm] at (image.west) {Frequency}; % Move "Frequency" closer
        \end{tikzpicture}
        \caption{google.com}
    \end{subfigure}
    % Second subfigure
    \begin{subfigure}[b]{0.32\textwidth}
        \centering
        \begin{tikzpicture}
            \node[anchor=south west,inner sep=0] (image) at (0,0) {\includegraphics[width=\textwidth]{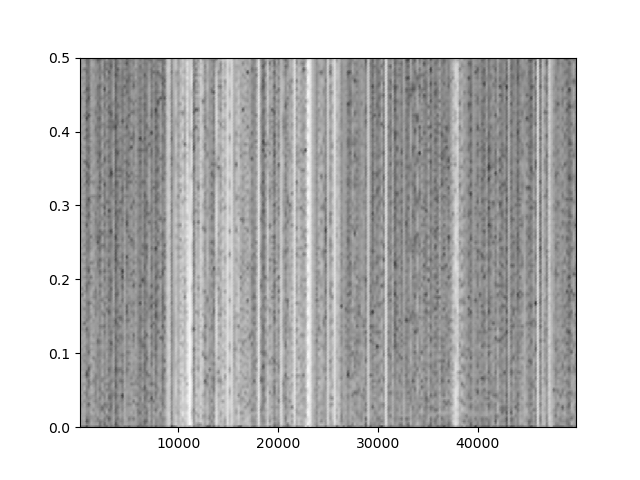}}; % Replace with your actual image file
            \node[above=-5mm] at (image.south) {Sample}; % Move "Sample" slightly farther away
            \node[rotate=90, anchor=south, yshift=-4mm] at (image.west) {Frequency}; % Move "Frequency" closer
        \end{tikzpicture}
        \caption{netflix.com}
    \end{subfigure}
    % Third subfigure
    \begin{subfigure}[b]{0.32\textwidth}
        \centering
        \begin{tikzpicture}
            \node[anchor=south west,inner sep=0] (image) at (0,0) {\includegraphics[width=\textwidth]{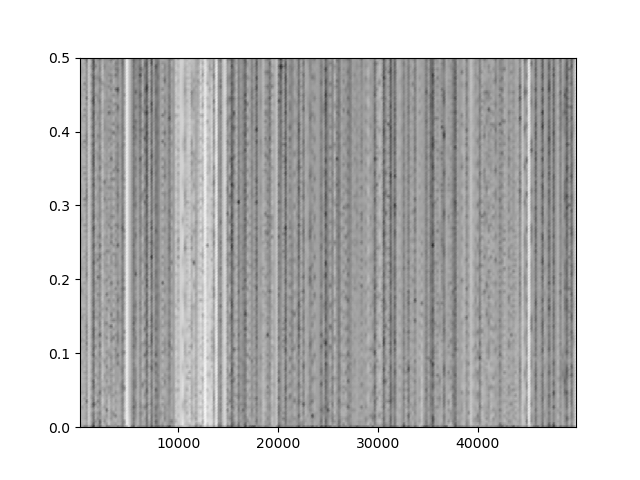}}; % Replace with your actual image file
            \node[above=-5mm] at (image.south) {Sample}; % Move "Sample" slightly farther away
            \node[rotate=90, anchor=south, yshift=-4mm] at (image.west) {Frequency}; % Move "Frequency" closer
        \end{tikzpicture}
        \caption{reddit.com}
    \end{subfigure}
    
    \caption{The STFTs of side-channel traces collected during the victim’s access to different websites show distinct patterns for (a) google.com, (b) netflix.com, and (c) reddit.com.}
    \label{fig:web_stft}
\end{figure*}
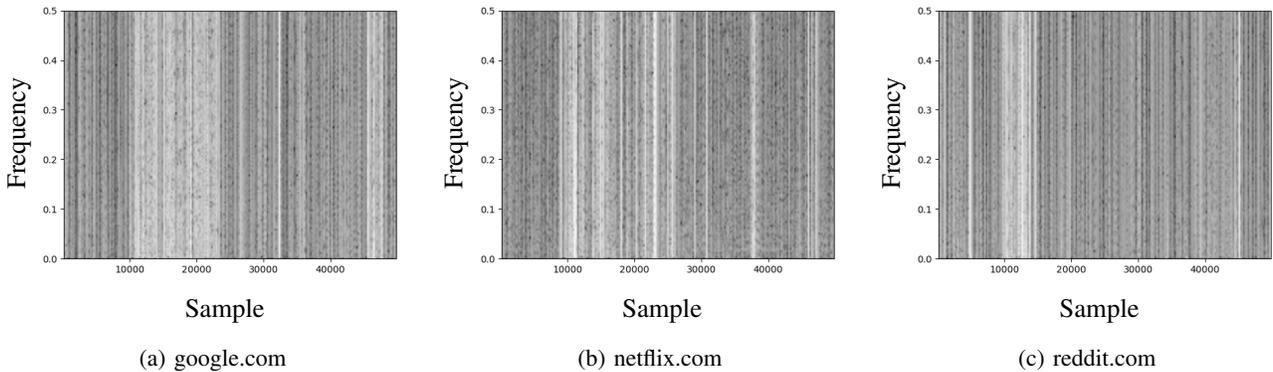

\subsection{Attack 1: Web Fingerprinting on Linux}
\label{subsec:web_finger}

In this attack, we demonstrate that an attacker can infer the specific website visited by the victim user in a closed-world setting over the top 100 websites from the Alexa top 1 million list~\cite{alexa_100} and an open-world setting with the top 100 websites and an \texttt{other-class} for websites that not listed in the top 100 websites. 

\textbf{{Experimental setup and attack overview.}} We run our experiments with the Chrome browser (Version 124.0.6367.201) on an i7-11800h CPU and Ubuntu 20.04. We assume the attacker is a normal user on that machine and can access the unprivileged \textit{rdtsc} timer and \textit{syncfs} system call. The system configurations are all in default with the ext4 file system. The attack and the victim co-locate on the main partition of the disk that is running the ext4 file system. We assume the victim Chrome browser has browser cache enabled as default. 

Our attack consists of two main stages: an online data collection phase and an offline phase for data preprocessing and trace evaluation. In the online phase, the attacker profiles the file system's write pattern in the background while the victim loads a website. As the website loads, the browser requests necessary files from the server and stores them in its cache. These temporary cache files are often partially encrypted and have limited user access. Although Chrome isolates each tab or process in a sandboxed environment to protect its cache, the browser cache files still share the page cache with other processes. Each website has a unique \textit{syncfs} delay pattern due to its specific write frequency and object sizes.

\textbf{Observing STFT distinguishability and classification.}
To process the data, we apply the short-time Fourier transform (STFT)~\cite{griffin1984signal} on the traces with a window size of 256 to capture frequency information. STFT highlights how frequency components change over time, refining signal analysis~\cite{rauscher2024idleleak, huang2019ecg_stft, chikkerur2007fingerprint_stft}. Using the \textit{scipy} package~\cite{virtanen2020scipy}, we implement STFT and present the results in Figure~\ref{fig:web_stft}, showing traces for \texttt{google.com}, \texttt{netflix.com}, and \texttt{reddit.com}. The x-axis represents sample numbers, with each trace containing 50,000 samples, while the y-axis indicates frequency changes. Observing the STFTs of these websites reveals clear distinguishability, enabling us to classify them by feeding these STFT images into a convolutional neural network (CNN).

We use the ResNet-152 CNN model~\cite{he2016deep} with pre-trained weights, customizing the first layer to accept a single input channel for our grayscale STFT images. The STFT images are resized to 224 by 224 pixels, and we train and evaluate the model in PyTorch~\cite{paszke2019pytorch} version 2.1.2 using 10-fold cross-validation~\cite{kohavi1995study}. For each fold, we evaluate performance using three metrics: F1 score (F1), Precision (Prec), and Recall (Rec). Our training setup utilizes the \textit{CrossEntropyLoss} function for multi-class classification and optimizes using the \textit{Adam} optimizer with a learning rate of 0.0001 and weight decay of 1e-5.

\textbf{{Evaluation in close-world setting.}} We collect 10,000 traces (100 per website) for a closed-world attack, resulting in 100 STFT images per website. The last layer of ResNet-152 is set to 100 classes, and the model is evaluated on an NVIDIA Tesla V100 GPU. We use a batch size of 80 for both training and evaluation. Table~\ref{tb:web_results} presents the performance of our side-channel attack on website fingerprinting. The average F1 score across 10 folds is 93.82\%, with a standard deviation of 6.96\%. Precision averages over 94\%, with a standard deviation of 5.58\%, while recall reaches an average of 93.77\%, with a 7.09\% standard deviation across 10-fold cross-validation.

 \textbf{{Evaluation in open-world setting.}} The classification results reported in previous sections were based on datasets collected in a closed-world setting (100 traces for each of the Alexa top 100 websites). To evaluate our attack in an open-world setting, we collected additional traces following the methodology outlined in prior works~\cite{ferguson2024webgpu, rauscher2024idleleak, wu2022rendering}. The open-world dataset includes the closed-world data plus 5,000 randomly selected Alexa websites (1 trace per website). We used the same ResNet-152 model, adjusting the last layer to classify 101 classes. Table~\ref{tb:web_results} presents our website fingerprinting attack's performance in the open-world setting. The average F1 score across 10 folds slightly decreases from 93.82\% to 93.25\%, with a standard deviation of 6.79\%, due to the addition of the new \texttt{other-class}. Precision remains above 94\%, with a standard deviation of 5.42\%, and recall averages 93.25\%, with a standard deviation of 6.67\% across 10-fold cross-validation.

% \begin{figure*}[ht]
%     \centering
%     % First row of figures
%     \subfigure[Reduction (from CUDA samples).]{
%         \includegraphics[width=0.31\textwidth]{figures/}
%         \label{fig:reduction}
%     }
%     \hfill
%     \subfigure[bf16TensorCoreGemm (from CUDA samples).]{
%         \includegraphics[width=0.31\textwidth]{figures/}
%         \label{fig:bf16}
%     }
%     \hfill
%     \subfigure[Srad (from Rodinia benchmarks).]{
%         \includegraphics[width=0.31\textwidth]{figures/}
%         \label{fig:srad}
%     }
%     \vspace{-0.3cm}
%     \caption{The STFTs of the traces of different website show distinct patterns: (a) and (b).}
%     \label{fig:web_stft}
% \end{figure*}

\begin{table}[]
\centering
% \normalsize
\caption{Web fingerprint performance: F1 (\%), Precision (\%), and Recall (\%).}
\begin{tabular}{c|c|c|c|}
\cline{2-4}
                                  & \textbf{F1}     & \textbf{Precision} & \textbf{Recall} \\ \cline{2-4} 
                                  & \textbf{$ \mu (\sigma)$} &  \textbf{$ \mu (\sigma)$}    & \textbf{$ \mu (\sigma)$} \\ \hline \hline
\multicolumn{1}{|c|}{Close world} & 93.82 (6.96)    & 94.99 (5.58)       & 93.77 (7.09)    \\ \hline
\multicolumn{1}{|c|}{Open world}  & 93.25 (6.79)               & 94.67 (5.42)                   &   93.25 (6.67)              \\ \hline
\end{tabular}
\label{tb:web_results}
\end{table}

\subsection{Attack 2: Video Fingerprinting on Linux}
\label{subsec:video_finger}

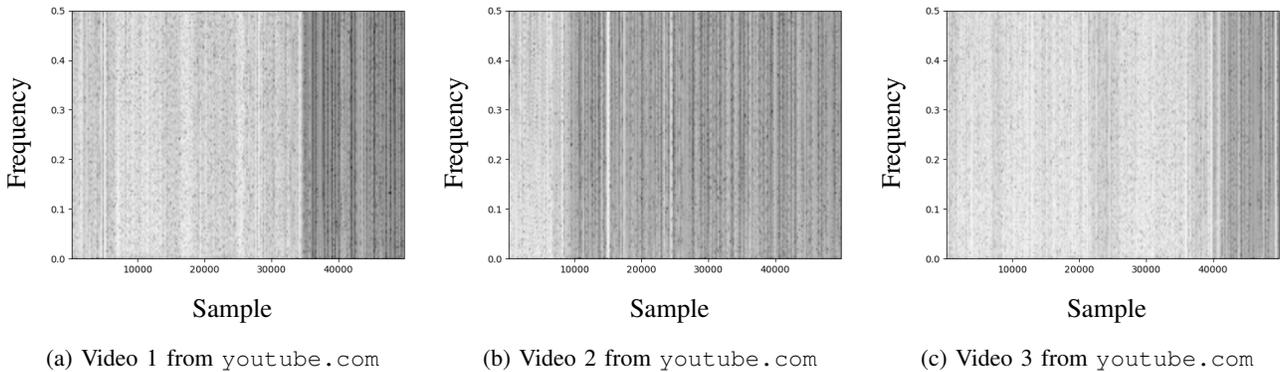
\begin{figure*}[ht]
    \centering
    % First subfigure
    \begin{subfigure}[b]{0.32\textwidth}
        \centering
        \begin{tikzpicture}
            \node[anchor=south west,inner sep=0] (image) at (0,0) {\includegraphics[width=\textwidth]{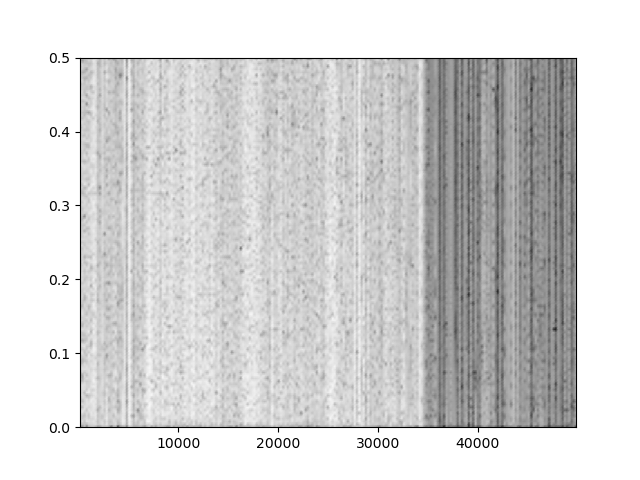}}; % Replace with your actual image file
            \node[above=-5mm] at (image.south) {Sample}; % Move "Sample" slightly farther away
            \node[rotate=90, anchor=south, yshift=-3mm] at (image.west) {Frequency}; % Move "Frequency" closer
        \end{tikzpicture}
        \caption{Video 1 from \texttt{youtube.com}}
    \end{subfigure}
    % Second subfigure
    \begin{subfigure}[b]{0.32\textwidth}
        \centering
        \begin{tikzpicture}
            \node[anchor=south west,inner sep=0] (image) at (0,0) {\includegraphics[width=\textwidth]{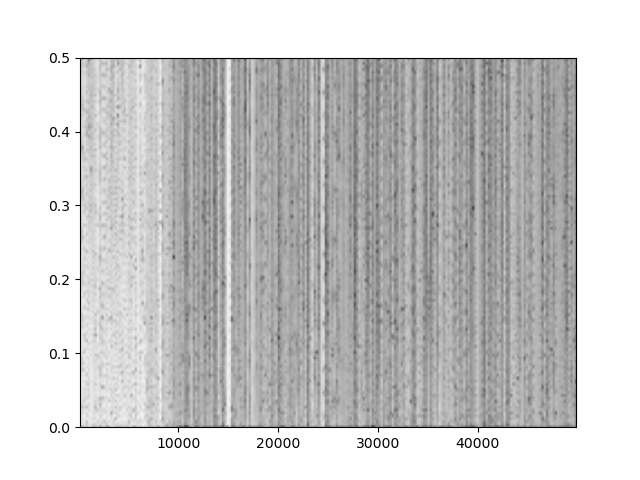}}; % Replace with your actual image file
            \node[above=-5mm] at (image.south) {Sample}; % Move "Sample" slightly farther away
            \node[rotate=90, anchor=south, yshift=-3mm] at (image.west) {Frequency}; % Move "Frequency" closer
        \end{tikzpicture}
        \caption{Video 2 from \texttt{youtube.com}}
    \end{subfigure}
    % Third subfigure
    \begin{subfigure}[b]{0.32\textwidth}
        \centering
        \begin{tikzpicture}
            \node[anchor=south west,inner sep=0] (image) at (0,0) {\includegraphics[width=\textwidth]{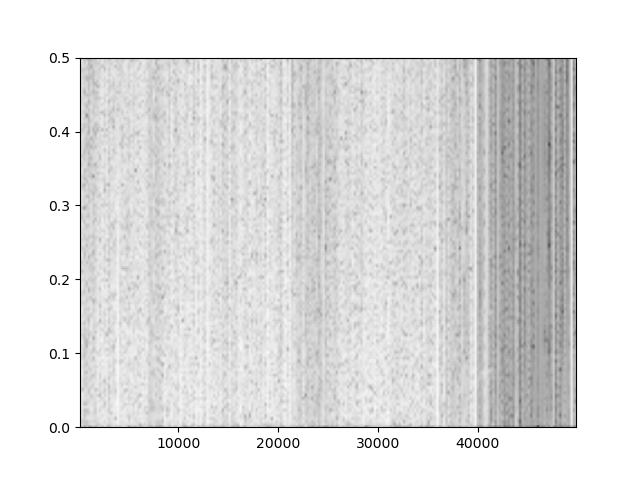}}; % Replace with your actual image file
            \node[above=-5mm] at (image.south) {Sample}; % Move "Sample" slightly farther away
            \node[rotate=90, anchor=south, yshift=-3mm] at (image.west) {Frequency}; % Move "Frequency" closer
        \end{tikzpicture}
        \caption{Video 3 from \texttt{youtube.com}}
    \end{subfigure}
    
    \caption{The STFTs of side-channel traces collected during the victim’s playback of different videos on \texttt{youtube.com} reveal distinct patterns for each of the three videos.}
    \label{fig:video_stft}
\end{figure*}

In this attack, we demonstrate that an attacker can infer the specific video played by the victim in a closed-world setting across the top 20 videos from two platforms: YouTube (\texttt{youtube.com}) and Bilibili (\texttt{bilibili.com}) trending lists. We also evaluate in an open-world setting, which includes the top 20 videos plus an \texttt{other-class} for videos not listed in the trending lists.

\textbf{{Experimental setup and data collection.}} In this attack, we conduct our experiments on an i7-11800h CPU with Ubuntu 20.04. We assume the attacker is a standard user on the machine with access to the unprivileged \textit{rdtsc} timer and \textit{syncfs} system call. Following the previous video fingerprinting attack setup, we use FFmpeg for live video streaming, as done in previous works~\cite{singh2018video, gu2018walls, garboan2012camcorder, blaskiewicz2020droppix}. FFmpeg, an open-source audio and video converter, supports most industry-standard codecs, enabling efficient file format conversion~\cite{tomar2006converting}. 

In our attack, we assume the victim uses \textit{ffplay}, a lightweight media player based on FFmpeg libraries, to watch video streams, while a separate host machine runs \textit{ffmpeg} as the streaming server. Meanwhile, the attacker profiles the write patterns of the victim's \textit{ffplay} in the background. For smooth playback, \textit{ffplay} uses a file system-backed temporary file to store received raw data. As the MPEG video is encoded, the received data is written to this file. The size and frequency of these writes are closely correlated with the visual content and dynamics of the video. Higher-resolution videos yield higher-quality signals, as the encoder operates more slowly and produces larger outputs, resulting in richer side-channel traces and improved classification accuracy. Consequently, the attacker can fingerprint the victim’s video by profiling these write patterns via the \textit{syncfs} system call.

 \textbf{Observing STFT Differentiability and Classification.} Following our web fingerprinting attack’s preprocessing method, we use STFT to extract frequency-based features from time-series leakages collected via the \textit{syncfs} system call. The STFT uses a window size of 256 to capture frequency information. Fig~\ref{fig:video_stft} shows the STFT images representing playback patterns for three videos when the victim streams them. The x-axis denotes sample numbers, with each trace containing 50,000 samples, while the y-axis represents frequency features. Observing these STFTs reveals clear differentiability, enabling the attacker to identify which video is playing on the victim’s device.

We use the same ResNet-152 CNN model as in our web fingerprinting attack, modifying the last layer to classify 20 classes. The model is evaluated using 10-fold cross-validation and three performance metrics: F1 score (F1), Precision (Prec), and Recall (Rec). Our training setup includes \textit{CrossEntropyLoss} as the loss function and \textit{AdamW} as the optimizer, with a learning rate of 0.0001 and weight decay of 1e-5.

\textbf{{Close-world setting.}} We collect 2,000 traces (100 per video) for a closed-world attack, resulting in 100 STFT images per video. A batch size of 20 is used for both training and evaluation. Table~\ref{tb:youtube_results} and Table~\ref{tb:bili_results} present the performance of our video fingerprinting attack across two platforms. On YouTube, the average F1 score across 10 folds exceeds 95\%, with a small standard deviation of 7.64\%. Precision averages 96.45\%, with a standard deviation of 6.22\%, while recall averages 95.68\%, with a 7.63\% standard deviation across 10-fold cross-validation. However, on Bilibili, the model performs slightly worse: the F1 score drops to 89.68\%, with a higher standard deviation of 15.84\%. Precision remains above 90\%, with a standard deviation of 14.57\%, and recall decreases to 89.80\%, with a standard deviation of 15.54\%. We conjecture that the difference in performance between the two platforms is due to video quality. On Bilibili, videos are played in low quality by default~\cite{zhang2023multi}, whereas YouTube does not impose this constraint. 

% The lower video quality of the Bilibili platform generates similar side channel leakages () makes our model harder to distingish the top 20 videos

\textbf{{Open-world setting.}} In addition to the top 20 videos in the closed-world setting, we add an \texttt{other-class} containing 50 randomly selected videos (1 trace per video). We retrained the same ResNet-152 model, adjusting the last layer to classify 21 classes. Table~\ref{tb:youtube_results} and Table~\ref{tb:bili_results} present the performance of our video fingerprinting attack in the open-world setting. On YouTube, the F1 score decreases by approximately 3\%, with precision dropping around 2\% and recall by 3\% compared to the closed-world setting. On Bilibili, performance also decreases, but the F1 score still exceeds 87\%, with precision reaching 89.26\% and recall remaining over 87\%.

\textbf{Web vs video.} We assume, consistent with prior web and video fingerprinting attacks, that the attacker already knows whether the victim is browsing the web or watching videos. However, this assumption is not critical since we observe that \textit{syncfs} exhibits distinct timing patterns across these activities that would allow an attacker to recognize when these applications execute. Thus, a spy could continuously collect \textit{syncfs} traces, which are then processed to identify when our target applications start to spy on them.  To provide evidence that the signal can be used to identify applications, we show that we can tell traces from our two target applications apart.  In particular, we train a binary classifier using the same CNN model employed in our fingerprinting attacks to infer the victim’s activity type, to provide evidence that the attack does not rely on knowing when the target applications are launched. The dataset contains 4,000 signal traces, evenly split between web browsing and video watching. Our model achieves an F1 score exceeding 99\% across 10-fold cross-validation, with a standard deviation of just 0.12\%. The average precision and recall are 99.98\% and 99.82\%, respectively, with standard deviations of 0.07\% and 0.25\%. %\nael{This classifier assumes that we know that one of them was launched and all we have to do is to tell them apart. I think the reviewer wants evidence that we can recognize when they get started at an arbitrary time or recognize them out of a group of other applications.  Perhaps we can rephrase the claim to something where this classifier supports the statement?}

\textbf{Robustness against noise.} Our preceding experiments were conducted in a relatively quiet environment. However, in real-world settings, other applications may also run on the same file system, introducing noise that degrades the quality of side-channel leakages. To evaluate the impact of such noise on the video fingerprinting attack, we ran a web browser in the background and opened additional browser tabs, then re-evaluated the attack under these conditions. We used the same experimental testbed and classifiers for consistency. As a result, when one browser tab was opened in the background, the average F1 score on YouTube dropped from 95.69\% to 79.57\%, with a standard deviation of 12.61\%. When five tabs were opened, the F1 score dropped further to 74.93\%, with a standard deviation of 13.75\%. Even with noise, the results remain far above the 5\% random-guess baseline for 20-class classification. This demonstrates that the side channel remains leaky and exploitable even in noisier environments.

\begin{table}[]
\centering
% \footnotesize
\caption{Video fingerprint performance on Youtube: F1 (\%), Precision (\%), and Recall (\%).}
\begin{tabular}{c|c|c|c|}
\cline{2-4}
                                  & \textbf{F1}     & \textbf{Precision} & \textbf{Recall} \\ \cline{2-4} 
                                  & \textbf{$ \mu (\sigma)$} & \textbf{$ \mu (\sigma)$}    & \textbf{$ \mu (\sigma)$} \\ \hline \hline
\multicolumn{1}{|c|}{Close world} &   95.69 (7.64)  &   96.45 (6.22)    &  95.68 (7.63) \\ \hline
\multicolumn{1}{|c|}{Open world}  &   92.74 (12.87)              &       94.70  (9.21)             &   92.69 (12.95)              \\ \hline
\end{tabular}
\label{tb:youtube_results}
\end{table}

\begin{table}[]
\centering
% \footnotesize
\caption{Video fingerprint performance on Bilibili: F1 (\%), Precision (\%), and Recall (\%).}
\begin{tabular}{c|c|c|c|}
\cline{2-4}
                                  & \textbf{F1}     & \textbf{Precision} & \textbf{Recall} \\ \cline{2-4} 
                                  & \textbf{$ \mu (\sigma)$} & \textbf{$ \mu (\sigma)$}    & \textbf{$ \mu (\sigma)$} \\ \hline \hline
\multicolumn{1}{|c|}{Close world} &  89.68 (15.84)  &   90.87 (14.57)     &   89.80 (15.54)  \\ \hline
\multicolumn{1}{|c|}{Open world}  &    87.03 (18.33)             &     89.26 (15.69)               &     87.26 (17.87)            \\ \hline
\end{tabular}
\label{tb:bili_results}
\end{table}

\subsection{Attack 3: Application Fingerprinting on Android}
\label{subsec:app_finger}

Information about other concurrent applications can reveal sensitive details about user activity and provide context for additional attacks, such as phishing~\cite{diao2016no} and keystroke inference~\cite{zhang2023s}. In this section, we show that an attacker can identify specific applications launched by the victim from a set of 15 popular Android applications.

\textbf{{Experimental setup and data collection.}} In this attack, we conduct our experiments on a Samsung S20 running Android 14. We assume the attacker is a standard user on this phone, with access to the unprivileged \textit{clock\_gettime} timer and \textit{syncfs} system call.  Following the setup of our previous fingerprinting attacks, we assume the victim launches applications in the foreground, while the attacker profiles the write patterns of these launches in the background. The sampling rate of the attacker's profiler is around 150,000 cycles/sec. Since each application launch generates unique write patterns in the Android system’s page cache, the attacker can infer the victim’s application by monitoring these patterns through the \textit{syncfs} system call. 
Table~\ref{tb:applist} lists the tested applications, all of which were downloaded from the Google Play Store. These apps represent the most popular choices within their respective categories. 

\begin{table}[tb]
\small
\centering
\begin{tabular}{|l|L{4.5cm}|}
\hline
\textbf{Category} & \textbf{Application}                                              \\ \hline
\hline
Social \& Entertainment      & Bluesky, Instagram, Tiktok, X, Netflix, IMDB, Youtube, Genshin Impact                          \\ \hline
Shopping \& E-commerce        & Amazon, Temu                    \\ \hline
Travel \& Services     & Airbnb, Uber \\ \hline
Finance \& Utilities     & Wells Fargo, ChatGPT, Google Map \\ \hline
\end{tabular}
\caption{Applications evaluated for the application fingerprinting attack.}
\label{tb:applist}
\end{table}

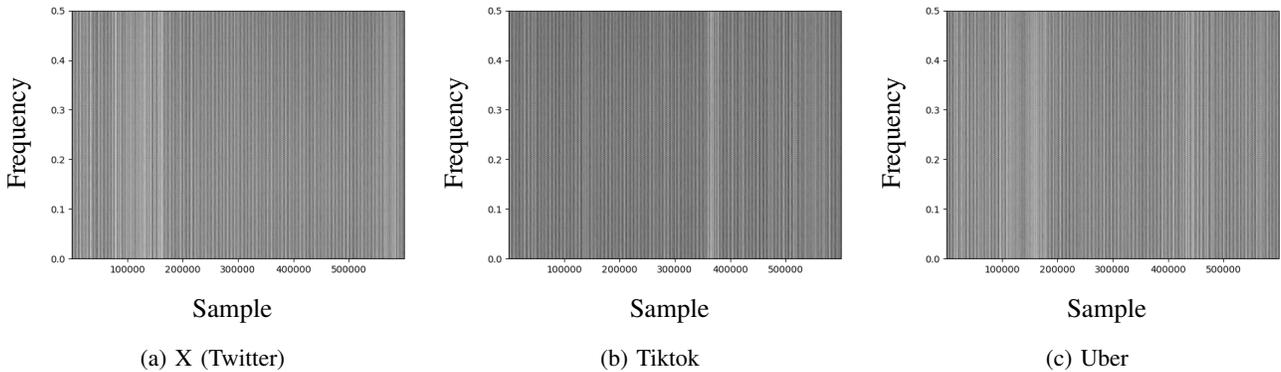
\begin{figure*}[t]
    \centering
    % First subfigure
    \begin{subfigure}[b]{0.32\textwidth}
        \centering
        \begin{tikzpicture}
            \node[anchor=south west,inner sep=0] (image) at (0,0) {\includegraphics[width=\textwidth]{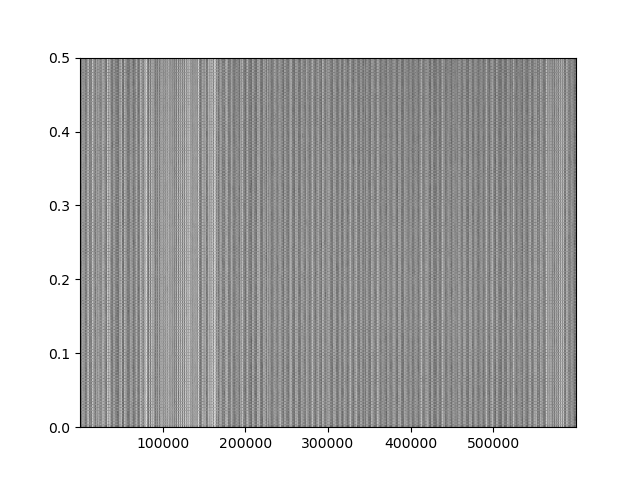}}; % Replace with your actual image file
            \node[above=-5mm] at (image.south) {Sample}; % Move "Sample" slightly farther away
            \node[rotate=90, anchor=south, yshift=-3mm] at (image.west) {Frequency}; % Move "Frequency" closer
        \end{tikzpicture}
        \caption{X (Twitter)}
    \end{subfigure}
    % Second subfigure
    \begin{subfigure}[b]{0.32\textwidth}
        \centering
        \begin{tikzpicture}
            \node[anchor=south west,inner sep=0] (image) at (0,0) {\includegraphics[width=\textwidth]{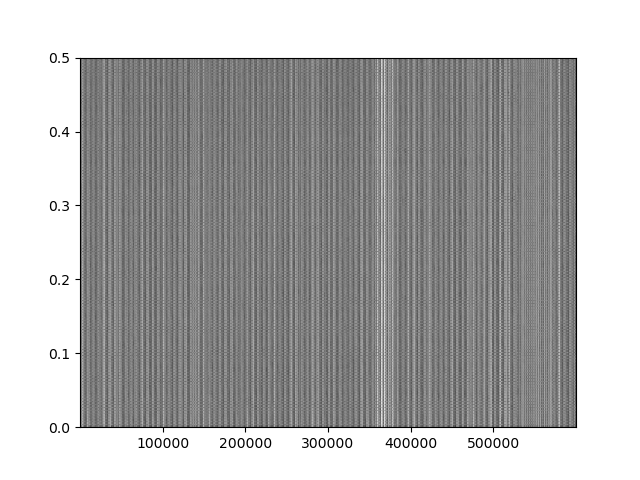}}; % Replace with your actual image file
            \node[above=-5mm] at (image.south) {Sample}; % Move "Sample" slightly farther away
            \node[rotate=90, anchor=south, yshift=-3mm] at (image.west) {Frequency}; % Move "Frequency" closer
        \end{tikzpicture}
        \caption{Tiktok}
    \end{subfigure}
    % Third subfigure
    \begin{subfigure}[b]{0.32\textwidth}
        \centering
        \begin{tikzpicture}
            \node[anchor=south west,inner sep=0] (image) at (0,0) {\includegraphics[width=\textwidth]{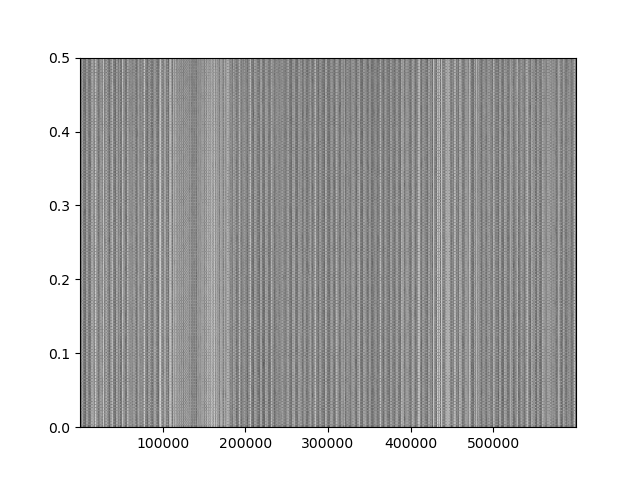}}; % Replace with your actual image file
            \node[above=-5mm] at (image.south) {Sample}; % Move "Sample" slightly farther away
            \node[rotate=90, anchor=south, yshift=-3mm] at (image.west) {Frequency}; % Move "Frequency" closer
        \end{tikzpicture}
        \caption{Uber}
    \end{subfigure}
    
    \caption{The STFTs of side-channel traces collected during the victim’s launch of different applications on an Android phone reveal distinct patterns for (a) X (Twitter), (b) TikTok, and (c) Uber.} 
    \label{fig:app_stft}
\end{figure*}

\textbf{Application fingerprinting demonstration and results.} We use STFT for data preprocessing to extract frequency-based features from time-series leakages collected via the \textit{syncfs} system call. The STFT uses a window size of 256 to capture frequency information. Fig~\ref{fig:app_stft} displays STFT images representing the launch patterns of three popular Android applications: X, TikTok, and Uber. The x-axis denotes sample numbers, with each trace containing 600,000 samples, while the y-axis represents frequency features. Utilizing the new timer on the Android phone provides a higher sample rate than in the previous two attacks. This results in clearly differentiated STFT images, allowing us to identify which application is concurrently running on the victim’s device. We use the same ResNet-152 CNN model as in our web fingerprinting attack, modifying the last layer to classify 15 classes. During training, we set \textit{CrossEntropyLoss} as the loss function and \textit{AdamW} as the optimizer, with a learning rate of 0.0001 and weight decay of 1e-5. The batch size for both training and testing is set to 80.

Table~\ref{tb:app_results} presents the performance of our application fingerprinting attack across 10 folds. The average F1 score is 93.49\%, with a standard deviation of 9.09\%. Precision averages 95.13\%, with a standard deviation of 6.85\%, while recall is over 96\%, with a standard deviation of 8.85\% across 10-fold cross-validation.

\begin{table}[t]
\centering
% \footnotesize
\caption{Application fingerprint performance on Android phone: F1 (\%), Precision (\%), and Recall (\%).}
\begin{tabular}{c|c|c|c|}
\cline{2-4}
                                  & \textbf{F1}     & \textbf{Precision} & \textbf{Recall} \\ \cline{2-4} 
                                  & \textbf{$ \mu (\sigma)$} & \textbf{$ \mu (\sigma)$}    & \textbf{$ \mu (\sigma)$} \\ \hline \hline
\multicolumn{1}{|c|}{Android} & 93.49 (9.09)  &   95.13 (6.85)     &   96.62 (8.85)  \\ \hline

\end{tabular}
\label{tb:app_results}
\end{table}

%% file: cross_container.tex
\section{Cross-container Attack}
% \yicheng{indetify when colocated container starts and ends. Add one figure to show starts and ends.}
Container technology is widely utilized by cloud computing service providers, such as Amazon Elastic Container Service (Amazon ECS)~\cite{AWS_docker} and Google Cloud Platform (GCP)~\cite{gcp_docker}. It offers a lightweight, operating system-level virtualization environment for efficient application hosting. In this attack, we investigate whether a spy can breach the isolation between two containers. In particular, we assume the spy resides in a separate container isolated from victim users, while these two containers share the same file system and hardware disks. The spy user continuously calls \textit{syncfs} in her own container while recording its time delay. When the victim process runs, any I/O system calls it performs that modify data or metadata increase the delay observed by the spy. The spy analyzes this delay pattern to infer the victim's file system activities. 

\textbf{Attack setup.} We conduct our experiments on a Linux machine equipped with an Intel i7-11750H CPU, using the NTFS file system with the NTFS3 driver. The NTFS file system was chosen for its better bandwidth performance compared to native Linux file system formats, as presented by our covert channel results in Table~\ref{tb:covert_results}. We use Docker Engine v20.10.17 to create two containers on the Linux machine.

\textbf{Experiment 4: I/O operations impacts on \textit{syncfs} execution in cross-container scenario.} In this experiment, we perform I/O operations in container A while measuring the execution time of \textit{syncfs} in container B. Each operation is repeated 1,000 times, and the average and standard deviation of the \textit{syncfs} execution time are calculated. Table~\ref{tb:latency_measurement_container} summarizes the \textit{syncfs} delay caused by four different I/O operations. Compared to the baseline, all I/O operations increase \textit{syncfs} delay by at least 3.1 times.

\begin{table}[tb]
\caption{\textit{syncfs} leakages for different I/O operations in cross-container attacks.}
\small
\begin{tabular}{|L{2.2cm}|C{1.9cm}|C{1.3cm}|C{1.5cm}|}

\hline
\textbf{I/O operation}         & \textbf{Affected buffers}              & \textbf{Average latency (cycles)} & \textbf{Standard deviation} \\ \hline \hline
baseline  & N/A                                    & 1526882                           & 243092                      \\ \hline
write                          & Page cache, journal and inode   & 19592860                        & 2617313                    \\ \hline
write(O\_SYNC)                 & journal                          & 4854316                          & 330761                      \\ \hline
ftruncate                      & journal and inode              & 10548983                          & 1026260                     \\ \hline
rename                         & journal and inode           & 10616821                          & 1729558                     \\ \hline
\end{tabular}
\label{tb:latency_measurement_container}

\end{table}

\begin{tcolorbox}[colback=gray!10,colframe=black,boxrule=0.5pt,arc=2mm,outer arc=2mm,]
\textit{\textbf{Observation 4}: I/O operations delay \textbf{syncfs} execution time in \textbf{cross-container scenarios}, with different types of operations causing varying levels of impact on \textbf{syncfs} latency.}
\end{tcolorbox}

\subsection{Detecting the Behavior of Container via \textit{syncfs}} 

\begin{figure}[tb]
    \centering
    \includegraphics[width=0.49\textwidth]{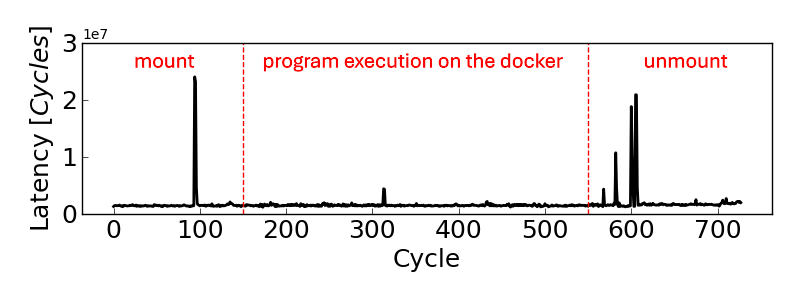}
    % \vspace{-2.0em}
    \caption{Detecting container startup and stop using \textit{syncfs}.}
    \label{fig: container detection}
\end{figure}

Containers must mount required directories during startup and unmount them when stopping, with these actions typically managed by the container runtime. In this section, we introduce a container detection technique leveraging \textit{syncfs}. The spy user profiles the latency pattern of \textit{syncfs} on container A while the victim user launches container B, mounts it to the file system, and runs a C program that sleeps for three seconds. Afterward, the victim stops container B and unmounts it from the file system. Figure~\ref{fig: container detection} presents leakage traces collected by the spy on container A. Two distinct spikes, corresponding to the mount and unmount operations, indicate container B's startup and shutdown, while the middle section represents the program execution.

Our detection method relies on lock contention, with the \textit{s\_umount} semaphore, a reader-writer lock in the superblock data structure, playing a key role. When the attacker and victim share the same disk partition, lock contention arises. In this attack, \textit{syncfs} acquires a reader lock during flushing, while mount and unmount operations acquire a writer lock. As a result, the spikes observed in the demo are caused by lock contention between these operations.

\subsection{Cross-container Covert Channel Design}

In a containerized environment, the sender and receiver are placed in separate containers, each managing a file within its working directory. We use the Timestamp Counter (TSC) as the synchronization method for the cross-container covert channel attack. Both the sender and receiver initialize a start TSC and begin transmitting messages once the current TSC matches the start TSC. The synchronization method is illustrated in Section~\ref{subsec:covert_design}.

To initiate the transmission, the sender transmits a predefined starting code to mark the beginning of the message. For data encoding, the sender uses direct I/O (\textit{write(O\_SYNC)}) to add a transaction to the journal cache to represent a bit "1" and executes a no-operation loop to represent a bit "0". We select \textit{write(O\_SYNC)} since its average latency is lowest among four operations as shown in Table~\ref{tb:latency_measurement_container}.

Simultaneously, the receiver repeatedly calls \textit{syncfs} in a loop, analyzing the \textit{syncfs} delay pattern to capture and decode the transmitted message. An ending code is appended to signify the termination of the message.

\subsection{Cross-container Covert Channel Evaluation} 
We then evaluate the cross-container covert channel in terms of bandwidth and error rate. In this attack, we achieve a bandwidth of 0.23 Kbps with an error rate of 2.4\%. The bandwidth of the cross-container covert channel is significantly lower compared to the standard covert channel described in Section~\ref{sec:covert_channel}. This reduction is attributed to the virtualization overhead introduced by Docker. For example, the baseline execution time of \textit{syncfs} in the cross-container scenario (when no other I/O operations are performed) is 1,526,882 cycles, which is over 600 times longer than in the same container scenario (2509 cycles as presented in Table~\ref{tb:latency_measurement}).

%% file: mitigation.tex
\section{Potential Mitigations} 
\label{sec:mitigation}

We propose four general classes of mitigations: (1) redesigning the \textit{syncfs} system call, (2) restricting access to high-resolution timers, (3) detecting suspicious system call queries, and (4) supporting isolation in I/O buffers. 
% These strategies align with mitigation approaches from prior side-channel attacks that exploit monitored states, such as \textit{procfs}~\cite{jana2012memento}, hardware interrupts~\cite{diao2016no}, and performance counters~\cite{zhang2023s}.

% \cheng{concern\#1}

\begin{figure}[tb]
    \centering
    \includegraphics[width=0.4\textwidth]{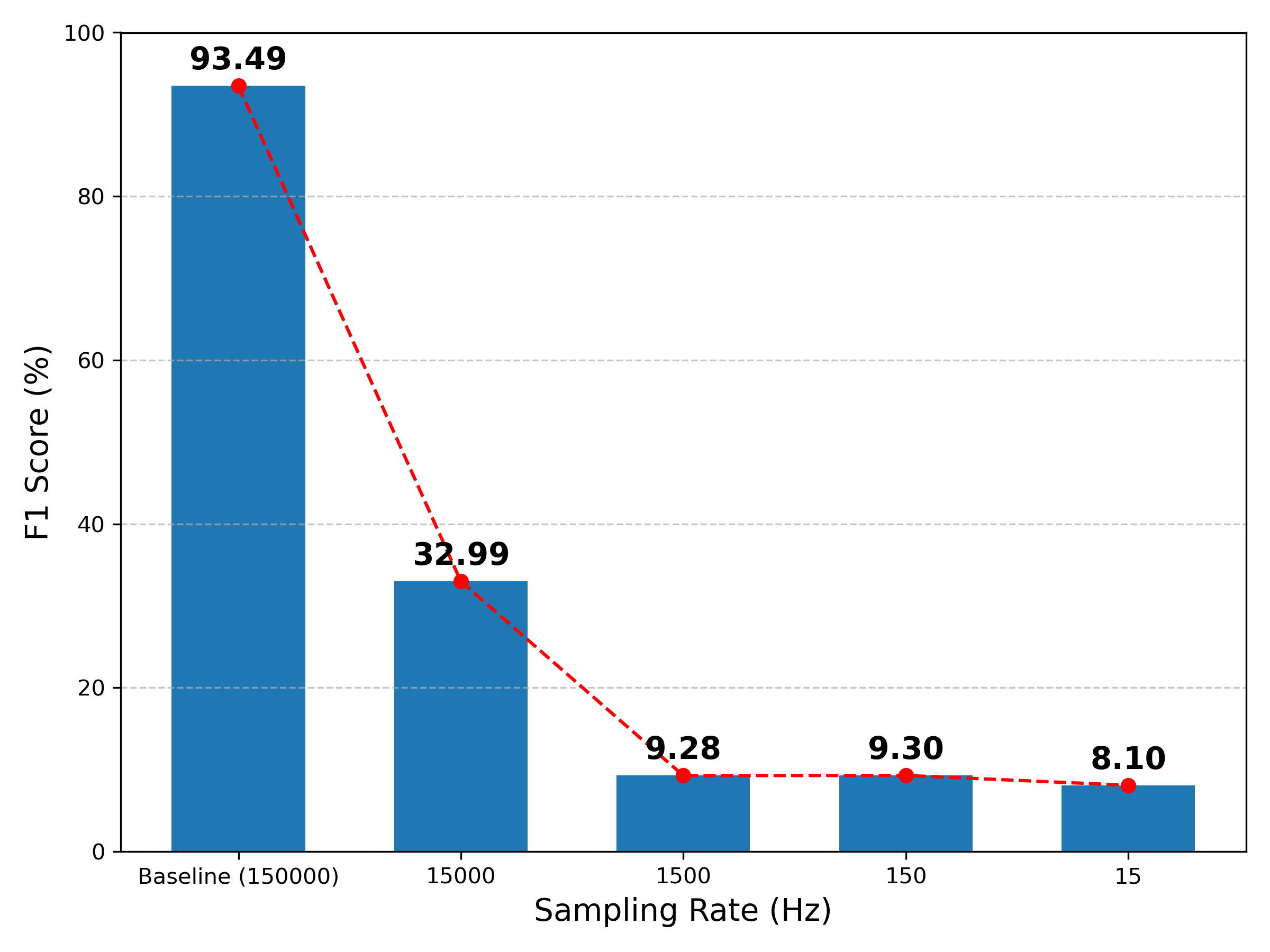}
    % \vspace{-2.0em}
    \caption{Application fingerprint attack results with limited frequency of \textit{syncfs} calls.}
    \label{fig:limiting_sampling}
\end{figure}

 \textbf{Redesigning \textit{syncfs}.} Completely blocking access to \textit{syncfs} could serve as an effective defense against our attacks. However, this would also disrupt legitimate users who depend on \textit{syncfs} to flush multiple file updates to disk. An alternative approach is to limit the frequency of \textit{syncfs} calls, which could diminish the effectiveness of spy applications~\cite{naghibijouybari2018rendered, wang2023danger}. Nonetheless, this strategy might impact legitimate users by potentially reducing \textit{syncfs} functionality and weakening persistence against system failures. Fig.~\ref{fig:limiting_sampling} shows the effect of reduced call frequency on application fingerprinting accuracy in Android.
We observe that reducing the sampling frequency from 150,000 to 15,000 Hz causes the attack's F1 score to drop significantly, from 93.49\% to 32.99\%. Further limiting the sampling rate to just 15 Hz reduces the attacker's F1 score to 8.1\%, which is nearly equivalent to random guessing (approximately 6.6\%, given the application fingerprinting involves 15 classes). However, since \textit{syncfs} plays a critical role in failure recovery, limiting the frequency of \textit{syncfs} queries may introduce drawbacks, such as an increased risk of data loss or a higher likelihood of file system inconsistencies~\cite{rebello2021can, verma2015failure}.

In addition, since \textit{syncfs} flushes the entire uncommitted page cache to the disk, a redesign to restrict users to flushing only their own page caches could mitigate the risk of cross-application data leakage. 
Another direction is to leverage hardware support, such as Physical Unclonable Functions (PUFs)~\cite{bai2016puf, bai2021novel} and Trusted Execution Environments (TEEs)~\cite{li2024sok}, to conceal timing leakages.

\textbf{Managing access to high-resolution clock instructions.} Our attacks rely on the high-resolution clock instructions, such as \textit{CNTVCT\_EL0} and \textit{rdtsc} timers, to measure the delay timing of \textit{syncfs}. Blocking access to these clock instructions could hinder such attacks; however, legitimate users also depend on them for precise program measurements. Additionally, even if certain clock instructions are restricted, attackers could exploit alternative low-level timing mechanisms, such as timer interrupts~\cite{schwarz2018keydrown}. Adding noise to these timers to obscure time differences has also been proposed as mitigation, as seen in prior work~\cite{zhang2024timing}. However, this approach incurs overhead, and recent studies~\cite{dutta2023spy} suggest that synthetic timers may still enable attackers to detect timing differences.

\textbf{Detecting abnormal \textit{syncfs} usage.} Our attack relies on continuous probing of \textit{syncfs} timing delays, which makes it a promising target for detection. A potential defense is to monitor abnormal \textit{syncfs} invocation patterns, such as periodic or high-frequency calls, by using a daemon process that tracks \textit{syncfs} activity. Prior research~\cite{alazab2020intelligent, salehi2017maar} has demonstrated the effectiveness of detecting malicious applications through API call sequence and return value monitoring, and similar techniques~\cite{ensemblehmd} could be applied here. 
Detection-based approaches may result in harmful false positives for applications that may naturally require higher frequencies of \textit{syncfs}.  Moreover, to evade such detection, an attacker may attempt to reduce the \textit{syncfs} query rate, trading stealth for reduced signal quality. As shown in Figure~{\ref{fig:limiting_sampling}}, we reduce the sampling rate from the maximum achievable values (150 kHz on Android and 120 kHz on Linux); this reduction comes at the cost of decreased attack success rate, but an attacker may still glean useful information while avoiding detection.  %Moreover, background activity and system noise often prevent \textit{syncfs} operation at these peak rates, further limiting the attacker's ability to maintain a high-resolution signal. However, some leakage may still persist even at lower rates, and setting too sensitive detection thresholds could lead to false positives.

\textbf{Supporting isolation in I/O buffers.} Without changing the functionality of \textit{syncfs}, we can redesign the write-back mechanism with minor changes. In our side channel attacks (§~\ref{subsec:web_finger} and §~\ref{subsec:video_finger}), we leak the write pattern of temporary files. These files do not need to be flushed to the disk since they should only be kept during the session. Therefore, a lightweight design is adding a tag for \textit{open} that prevents the dirty pages of this file from being flushed by \textit{syncfs}. We leave this as future work.

%% file: relatedwork.tex
\section{Related Work}
\label{sec:relatedwork}

% \yicheng{Todo}
\textbf{OS-level covert and side channel attacks.} The OS is designed to ensure isolation between multiple users or processes. However, researchers have demonstrated that side-channel leakages can compromise this isolation, exposing sensitive information like passwords and keystrokes. On Linux, Zhang and Wang~\cite{zhang2009peeping} first demonstrated side-channel attacks via \textit{procfs}, enabling the eavesdropping of users' keystrokes. Qian et al.~\cite{qian2012collaborative} then leveraged error packet counters in \texttt{/proc/net/} to execute off-path TCP session hijacking. Recently, Shen et al.~\cite{shen2023mes} exploited mutual exclusion and synchronization mechanisms (MES) to establish covert channels for data transmission.
On Android, 
% Chen et al.~\cite{chen2014peeking} revealed a shared-memory side channel in the GUI framework that allows UI state inference without requiring permissions. 
Diao et al.~\cite{diao2016no} identified critical information leakage in Android's interrupt handling mechanism. Similarly, on iOS, Zhang et al.~\cite{zhang2018level} were the first to uncover OS-level side-channel vulnerabilities in iOS, exposing attack vectors such as memory and network usage APIs. Our work uncovers a novel vector of OS-level leakage, focusing specifically on file system synchronization calls within shared file systems.

\textbf{I/O-related attacks.}  Our attacks fall into the category of I/O-related attacks. Prior research has demonstrated that contention leakages on I/O can be exploited to establish covert and side channels. On one hand, timing leakages from contention have been shown on physical I/O components in modern computers, such as PCIe, CPU ring bus, and NVLink. For example, Tan et al.~\cite{tan2021invisible} introduced PCIe congestion side-channel attacks across GPUs, Network Interface Cards (NICs), and Solid-State Drives (SSDs). Similarly, Mert Side et al.~\cite{side2022lockeddown} introduce a novel side channel vulnerability exploiting contention on the host-GPU PCIe bus, a critical interface in modern GPUs used in both traditional and cloud computing.  Paccagnella et al.~\cite{paccagnella2021lord} and Dutta et al.~\cite{dutta2021} propose microarchitectural side channel attacks leveraging contention on the CPU ring interconnect. Similar contention-based attacks have also been executed on other hardware architectures, such as CPU Mesh~\cite{wan2022meshup}. 
% Wan et al.~\cite{wan2022meshup} present MESHUP, a stateless cache side-channel attack targeting the mesh interconnect of server-grade CPUs. 
% Dai et al.~\cite{dai2022don} also demonstrate side-channel attacks targeting Intel's mesh interconnect, revealing how attackers can exploit contention to build a 1.53 Mbps covert channel and extract keys from cryptographic implementations. 
Recently, Zhang et al.~\cite{zhang2024beyond} reveal vulnerabilities in NVLink interconnects, demonstrating a covert channel across GPUs and an application fingerprinting side channel attack using congestion-based timing leakages. 
On the other hand, software I/O has also been investigated for contention-based leakages. Gruss et al.~\cite{gruss2019page} introduced a software side-channel attack on the OS page cache, allowing unprivileged monitoring of memory access patterns. 

\textbf{Comparison with \textit{fsync}-based attacks.}
More recently, Chen et al.{~\cite{chen2023sync+}} proposed a set of software cache write covert channels that exploit file system synchronization primitives. Similarly, Jiang et al.{~\cite{jiang2024sync+}} presented Sync+Sync, a timing-based covert channel leveraging contention in \textit{fsync} calls on shared persistent storage. Both of these works depend on contention from concurrent \textit{fsync} API invocations and are limited to forming covert channels. 
Jiang et al.{~\cite{jiang2024sync+}} introduced \textit{fsync}-based contention leakage for website fingerprinting, requiring both the victim and attacker to use the \textit{fsync} system call to create contention. However, their approach is limited by the similar number of \textit{fsync} calls across websites, resulting in low accuracy for website fingerprinting, with reliable identification for \textbf{only two} websites (\texttt{qq.com} and \texttt{sina.com.cn}). 
In contrast, our method utilizes each website's unique \textit{syncfs} delay pattern, which reflects the sequence of write operations and sizes, patterns that are unique to the website content and structure.  As a result, the attack achieves over 93\% average F1 score in a 100-class website fingerprinting attack.

% \hlc[cyan!40]{Prior work on web fingerprinting using \textit{fsync} is limited by the similar number of \textit{fsync} calls across websites, leading to low classification accuracy. In fact, reliable identification was achieved for only two websites: \texttt{qq.com} and \texttt{sina.com.cn}. In contrast, our approach leverages each website’s unique \textit{syncfs} delay pattern, which captures differences in write frequency and content size. It requires no special behavior from the victim (such as explicitly invoking \textit{fsync}) and achieves over 93\% average F1 score in a 100-class website fingerprinting task.
% }
% Our work exploits delay patterns in \textit{syncfs}, a fundamentally different approach to contention-based leakages, which does not require the victim application to execute any I/O system calls, enabling us to implement both robust covert channels and high-resolution side channel attacks.

\textbf{Cross-container attacks.} Multi-tenancy is the feature of cloud computing that allows computation instances from different tenants to run on the same physical server. However, a series of work has demonstrated the cross-container attack on the cloud.  
 Ristenpart et al.~\cite{ristenpart2009hey}  is the first work on co-residence detection on the commercial cloud server. Gao et al.~\cite{gao2018study} identify information leakage channels in multi-tenancy container clouds caused by incomplete Linux kernel isolation mechanisms. Recently, Zhao et al.~\cite{zhao2024last} conducted co-location attacks in public Function-as-a-Service (FaaS) environments, focusing on Google Cloud Run. Later, they successfully an end-to-end LLC Prime+Probe side-channel attack on Google Cloud Run~\cite{zhao2024everywhere}.

%% file: conclusion.tex
\section{Concluding Remarks}

This work identified vulnerabilities in the \textit{syncfs} system call that expose timing-based leakage vectors, compromising logical isolation. By exploiting these weaknesses, we demonstrated high-bandwidth covert channels. Additionally, we developed three side-channel attacks for fingerprinting websites, videos, and applications. Furthermore, we explored cross-container attacks that breach isolation in containerized environments. Our findings highlight the need for improved defenses, such as redesigning \textit{syncfs} and enhancing I/O buffer isolation. These measures are crucial to mitigating the identified risks while maintaining system functionality.
Our attacks demonstrate that \textit{syncfs}-based vulnerabilities pose a significant threat to system security.

\section*{Acknowledgements}
This research was sponsored in part by the OUSD(R\&E)/RT\&L and was accomplished under Cooperative Agreement Number W911NF-20-2-0267. The views and conclusions contained in this document are those of the authors and should not be interpreted as representing the official policies, either expressed or implied, of the ARL and OUSD(R\&E)/RT\&L or the U.S. Government. The U.S. Government is authorized to reproduce and distribute reprints for Government purposes notwithstanding any copyright notation herein.  We would like to thank the paper reviewer and, especially, the shepherd whose comments and requests substantially improved the paper.

% In this paper, we present covert and side-channel attacks exploiting a novel attack vector based on \textit{syncfs}. By measuring the delay patterns of \textit{syncfs}, we can identify different I/O system call footprints and write sizes. Leveraging these insights, we design a \textit{syncfs} covert channel attack and evaluate its effectiveness on both Linux native file systems and the Windows file system. Additionally, we develop three end-to-end side-channel attacks targeting Linux and Android environments.

% Furthermore, we explore cross-container attacks using \textit{syncfs}, including a container detection technique and a cross-container covert channel attack. Our findings expose a significant vulnerability in \textit{syncfs} and Linux I/O buffers, highlighting the need for enhanced security measures.

% In this paper, we demonstrate covert and side channel attacks using a new attack vector, \textit{syncfs}. In reverse engineering, we identify two leakage vectors through \textit{syncfs}: I/O system call footprint and write size. We utilize these observations to construct a \textit{syncfs} covert channel attack and evaluate it on both Linux native file systems and the Windows file system. We also build three end-to-end side channel attacks on Linux and Android. Additionaly, we present cross-container attacks using \textit{syncfs} including a container detection technique and a cross-container covert channel attack. Our work reveals the a significant vulnerability in \textit{syncfs} and Linux I/O buffers.    

%% file: meta_review.tex
\newpage % The Meta-Review should at least start on a new column

% Use \appendices and not \appendix due to IEEEtran.cls quirks
\appendices % if not used earlier

\section{Meta-Review}

\subsection{Summary}
This paper introduces a new class of side-channel attacks exploiting shared file system code. It identifies the syncfs system call as a new source of leakage that enables an attacker to infer activities of other processes on the system.

\subsection{Scientific Contributions}

\begin{itemize}
\item Identifies an Impactful Vulnerability
\item Provides a Valuable Step Forward in an Established Field
\end{itemize}

\subsection{Reasons for Acceptance}
\begin{enumerate}
\item This paper identifies an impactful vulnerability in file system code that is present in virtually all existing Linux systems. It demonstrates how to carefully control and leverage syncfs delays to instantiate covert channels and infer behaviors of victim processes.
\item The paper provides a valuable step forward in an established field. Secure file and storage systems have been well-studied, but recent security research has shown that side-channel attacks are present at virtually all layers of the software stack. This paper highlights that even core system software is vulnerable to side-channel attacks and provides some insights on potential mitigations.
\end{enumerate}

\subsection{Noteworthy Concerns} % Exclude if your meta-review does not have noteworthy concerns
\begin{enumerate} % Enumerate environment is not necessary if there is only one
\item  The attacks rely on high sampling rates (150K calls per second). Attacks therefore have anomalous behavior that might be easily detected by simple (or state of the art) anomaly detection techniques. The takeaways could be stronger if the paper evaluated the detectability of the attack to some extent.
\end{enumerate}

\section{Response to the Meta-Review} % Optional

We thank the reviewers and the shepherd for their insightful comments. Current defenses do not detect the attack.  We acknowledge that anomaly detection mechanisms might be configured to flag high \texttt{syncfs()} rates. However, it is not clear whether a low detection threshold is feasible without introducing false positives or whether the attacker can adjust the sampling rate to remain below the detection threshold while still exploiting the side channel.